\documentclass{article}
\usepackage{a4}
\usepackage[T1]{fontenc}
\usepackage{times}
\usepackage{mathptm}
\usepackage{amsmath}
\usepackage{amssymb}
\usepackage{amsthm}
\usepackage{exscale}

\makeatletter
\renewcommand{\@endtheorem}{\endtrivlist}
\makeatother

\numberwithin{equation}{section}

\newcommand{\Cbb}{\mathbb{C}}
\newcommand{\Nbb}{\mathbb{N}}
\newcommand{\Rbb}{\mathbb{R}}

\newcommand{\Wecm}{\mathecm{W}}

\newcommand{\Afrak}{\mathfrak{A}}
\newcommand{\Cfrak}{\mathfrak{C}}
\newcommand{\Lfrak}{\mathfrak{L}}
\newcommand{\Nfrak}{\mathfrak{N}}
\newcommand{\Tfrak}{\mathfrak{T}}

\newcommand{\zeroib}{\mathib{0}}
\newcommand{\Kib}{\mathib{K}}
\newcommand{\Oib}{\mathib{O}}
\newcommand{\hib}{\mathib{h}}
\newcommand{\pib}{\mathib{p}}
\newcommand{\vib}{\mathib{v}}
\newcommand{\xib}{\mathib{x}}
\newcommand{\yib}{\mathib{y}}
\newcommand{\zib}{\mathib{z}}

\newcommand{\Hscr}{\mathscr{H}}
\newcommand{\Oscr}{\mathscr{O}}
\newcommand{\Sscr}{\mathscr{S}}

\newcommand{\Ssf}{\mathsf{S}}
\newcommand{\Vsf}{\mathsf{V}}

\newcommand{\unit}{\mathbf{1}}

\newcommand{\Rs}{\mathbb{R}^s}
\newcommand{\Rsone}{\mathbb{R}^{s + 1}}

\newcommand{\AS}{\mathfrak{A}_{\mathscr{S}}}
\newcommand{\Aann}{\mathfrak{A}_{\text{\itshape ann}}}

\newcommand{\Deltabar}{\overline{\Delta}}

\newcommand{\Deltaetabar}{\overline{\Delta}_\eta}

\newcommand{\BH}{\mathscr{B} ( \mathscr{H} )}

\newcommand{\BHw}{\mathscr{B} ( \mathscr{H}_w )}

\newcommand{\Lor}{\mathsf{L}_+^\uparrow}
\newcommand{\Poin}{\mathsf{P}_{\negthinspace +}^\uparrow}

\newcommand{\fwcone}{\overline{V}_{\negthinspace +}}

\newcommand{\ED}{E ( \Delta )}
\newcommand{\EDprime}{E ( \Delta' )}
\newcommand{\EDbar}{E ( \overline{\Delta} )}
\newcommand{\EDetabar}{E ( \overline{\Delta}_\eta )}
\newcommand{\EsDprime}{E_\sigma ( \Delta' )}
\newcommand{\EsDprimeeta}{E_\sigma ( \Delta'_\eta )}
\newcommand{\EsGamma}{E_\sigma ( \Gamma )}
\newcommand{\EwDprime}{E_w ( \Delta' )}
\newcommand{\EwGamma}{E_w ( \Gamma )}

\newcommand{\Cstar}{\mathfrak{C}^*}
\newcommand{\Cstarplus}{\mathfrak{C}^{* \thinspace +}}

\newcommand{\CDstar}{{\mathfrak{C}_\Delta}^{\negthickspace *}}
\newcommand{\CDstarplus}{{\mathfrak{C}_\Delta}^{\negthickspace *
    \thinspace +}}

\newcommand{\aLax}{\alpha_{( \Lambda , x )}}

\newcommand{\agGamma}{\alpha_{g_\Gamma}}
\newcommand{\aibx}{\alpha_\mathib{x}}
\newcommand{\aibxprime}{\alpha_{\mathib{x}'}}
\newcommand{\aibtwox}{\alpha_{2\mathib{x}}}
\newcommand{\aiby}{\alpha_\mathib{y}}
\newcommand{\aminusibx}{\alpha_{(-\mathib{x})}}
\newcommand{\ax}{\alpha_x}

\newcommand{\AOr}{\mathfrak{A} ( \mathscr{O}_r )}

\newcounter{defitem}

\newenvironment{deflist}{\begin{list}{(\alph{defitem})}%
  {\usecounter{defitem} \setlength{\topsep}{0ex}%
   \setlength{\parsep}{0.2ex} \setlength{\itemsep}{0.4ex}%
   \setlength{\leftmargin}{0em} \setlength{\itemindent}{0.5em}%
   }}{\end{list}}

\newcounter{proofitem}

\newenvironment{prooflist}{\begin{list}{(\roman{proofitem})}%
  {\usecounter{proofitem} \setlength{\topsep}{0ex}%
   \setlength{\parsep}{0.2ex} \setlength{\itemsep}{0.4ex}%
   \setlength{\leftmargin}{0em} \setlength{\itemindent}{0.5em}%
   \setlength{\listparindent}{1em}}}{\qed \end{list}}

\newcounter{Proofitem}

\newenvironment{Prooflist}{\begin{list}{Part (\Roman{Proofitem}):}%
  {\usecounter{Proofitem} \setlength{\topsep}{0ex}%
   \setlength{\parsep}{0.2ex} \setlength{\itemsep}{0.4ex}%
   \setlength{\leftmargin}{0em} \setlength{\itemindent}{0.5em}%
   \setlength{\listparindent}{1em}}}{\qed \end{list}}

\newcounter{Proofsubitem}

\newcounter{propitem}

\newenvironment{proplist}{\begin{list}{(\roman{propitem})}%
  {\usecounter{propitem} \setlength{\topsep}{0ex}%
   \setlength{\parsep}{0.2ex} \setlength{\itemsep}{0.4ex}%
   \setlength{\leftmargin}{0em} \setlength{\itemindent}{0.5em}%
   }}{\end{list}}

\newcounter{remitem}

\newenvironment{remlist}{\begin{list}{(\roman{remitem})}%
  {\usecounter{remitem} \setlength{\topsep}{0ex}%
   \setlength{\parsep}{0.2ex} \setlength{\itemsep}{0.4ex}%
   \setlength{\leftmargin}{0em} \setlength{\itemindent}{0.5em}%
   }}{\end{list}}

\newcounter{theoitem}

\newenvironment{theolist}{\begin{list}{(\Roman{theoitem})}%
  {\usecounter{theoitem} \setlength{\topsep}{0ex}%
   \setlength{\parsep}{0.2ex} \setlength{\itemsep}{0.4ex}%
   \setlength{\leftmargin}{0em} \setlength{\itemindent}{0.5em}%
   }}{\end{list}}

\newcounter{theosubitem}

\newenvironment{theosublist}{\begin{list}{(\roman{theosubitem})}%
  {\usecounter{theosubitem} \setlength{\topsep}{0ex}%
   \setlength{\parsep}{0.2ex} \setlength{\itemsep}{0.4ex}%
   \setlength{\leftmargin}{0em} \setlength{\itemindent}{2em}%
   }}{\end{list}}

\theoremstyle{definition}
\newtheorem{definition}{Definition}[section]

\theoremstyle{plain}
\newtheorem{theorem}[definition]{Theorem}
\newtheorem{proposition}[definition]{Proposition}
\newtheorem{corollary}[definition]{Corollary}
\newtheorem{lemma}[definition]{Lemma}

\theoremstyle{remark}
\newtheorem*{remark*}{Remark}

\DeclareMathAlphabet{\mathib}{T1}{ptm}{b}{it}
\DeclareMathAlphabet{\mathcmmib}{OML}{cmm}{b}{it}
\DeclareMathAlphabet{\mathecm}{U}{eur}{m}{n}
\DeclareMathAlphabet{\mathscr}{U}{rsfs}{m}{n}

\DeclareMathOperator{\linhull}{span}

\DeclareMathOperator{\supp}{supp}
\DeclareMathOperator{\sw}{\sigma-weak}

\newcommand{\set}[1]{\{ #1 \}}
\newcommand{\bset}[1]{\bigl\{ #1 \bigr\}}
\newcommand{\Bset}[1]{\Bigl\{ #1 \Bigr\}}

\newcommand{\bcomm}[2]{\bigl[ #1 , #2 \bigr]}
\newcommand{\Bcomm}[2]{\Bigl[ #1 , #2 \Bigr]}

\newcommand{\abs}[1]{\lvert #1 \rvert}
\newcommand{\babs}[1]{\bigl\lvert #1 \bigr\rvert}
\newcommand{\Babs}[1]{\Bigl\lvert #1 \Bigr\rvert}

\newcommand{\norm}[1]{\lVert #1 \rVert}
\newcommand{\bnorm}[1]{\bigl\lVert #1 \bigr\rVert}
\newcommand{\Bnorm}[1]{\Bigl\lVert #1 \Bigr\rVert}

\newcommand{\pD}{p_\Delta}
\newcommand{\pDx}[1]{p_\Delta ( #1 )}
\newcommand{\bpDx}[1]{p_\Delta \bigl( #1 \bigr)}
\newcommand{\BpDx}[1]{p_\Delta \Bigl( #1 \Bigr)}

\newcommand{\pDprimex}[1]{p_{\Delta'} ( #1 )}

\newcommand{\qD}{q_\Delta}
\newcommand{\qDx}[1]{q_\Delta ( #1 )}
\newcommand{\bqDx}[1]{q_\Delta \bigl( #1 \bigr)}
\newcommand{\BqDx}[1]{q_\Delta \Bigl( #1 \Bigr)}

\newcommand{\qDprimex}[1]{q_{\Delta'} ( #1 )}

\newcommand{\qs}{q_\varsigma}
\newcommand{\qsx}[1]{q_\varsigma ( #1 )}

\newcommand{\qw}{q_w}
\newcommand{\qwx}[1]{q_w ( #1 )}

\newcommand{\ket}[1]{\vert #1 \rangle}
\newcommand{\bket}[1]{\big\vert #1 \bigr\rangle}

\newcommand{\scp}[2]{\langle #1 \vert #2 \rangle}
\newcommand{\bscp}[2]{\bigl\langle #1 \big\vert #2 \bigr\rangle}

\newcommand{\scpx}[3]{\langle #1 \vert #2 \vert #3 \rangle}
\newcommand{\bscpx}[3]{\bigl\langle #1 \big\vert #2 \big\vert #3
  \bigr\rangle}

\begin{document}
\title{Particle Weights and their Disintegration I}
\author{Martin Porrmann\\
  II. Institut f\"ur Theoretische Physik, Universit\"at Hamburg\\
  Luruper Chaussee 149, D-22761 Hamburg, Germany\\
  e-mail: \texttt{martin.porrmann@desy.de}}
\date{November 6, 2002}

\maketitle

\begin{abstract}
  The notion of Wigner particles is attached to irreducible unitary
  representations of the Poincar\'e group, characterized by parameters
  $m$ and $s$ of mass and spin, respectively. However, the Lorentz
  symmetry is broken in theories with long-range interactions,
  rendering this approach inapplicable (infraparticle problem). A
  unified treatment of both particles and infraparticles via the
  concept of particle weights can be given within the framework of
  local quantum physics. They arise as temporal limits of physical
  states in the vacuum sector and describe the asymptotic particle
  content. In this paper their definition and characteristic
  properties, already presented in \cite{buchholz/porrmann/stein:1991}
  and \cite{haag:1996}, are worked out in detail. The existence of the
  temporal limits is established by use of suitably defined seminorms
  which are also essential in proving the characteristic features of
  particle weights.
\end{abstract}

\section{Introduction}
  \label{sec:introduction}
  
  Physical phenomena in high energy physics are analyzed in terms of
  particles arising as asymptotic configurations of elementary
  localized entities in scattering experiments. These particles are
  characterized by certain intrinsic properties expressed by quantum
  numbers; their consistent and complete theoretical explanation is a
  goal of quantum field theory. The usual theoretical description of
  particles goes back to Wigner's classical analysis of irreducible
  unitary representations of the Poincar\'e group \cite{wigner:1939}.
  He gives a complete classification of these representations labelled
  by two parameters $m$ and $s$. It is standard to assume that states
  of an elementary particle pertain to one of these, interpreting the
  corresponding parameters as the intrinsic mass and spin,
  respectively, of the particle. However, this approach is not
  universally applicable, for there are quantum field theoretic models
  allowing for particles of zero rest mass, where states of particles
  coupled to these cannot be described in terms of eigenstates of the
  mass operator. An example is quantum electrodynamics, where charged
  particles are inevitably accompanied by soft photons. It is an open
  question, known as the infraparticle problem \cite{schroer:1963},
  how mass and spin are to be described in this situation. Moreover,
  standard collision theory does not work either. The present paper
  describes an alternative approach to the concept of particles within
  the framework of local quantum physics
  \cite{haag/kastler:1964,haag:1996}. The fundamental ideas and
  results are due to Buchholz and have first been presented in
  \cite{buchholz:1986b}. Concepts and statements may be familiar from
  \cite{buchholz/porrmann/stein:1991} (cf.~also
  \cite[Section~VI.2.2]{haag:1996}); the more detailed elaboration
  below is the result of joint work of Detlev Buchholz and the author.
  
  The particle concept to be set forth is motivated by the
  experimental situation encountered in high energy physics, where
  certain entities show up as \emph{particles}, being traced by
  specific measuring devices called \emph{detectors}. The invariant
  characteristic of these physical systems in different circumstances
  is their being localized in the interaction with a detector. In 1967
  Araki and Haag \cite{araki/haag:1967} presented an analysis of the
  particle content of scattering states for \emph{massive} models,
  investigating the asymptotics of physical states by means of certain
  operators to be interpreted as counters. These are assumed to be
  well localized and, at the same time, to annihilate the vacuum in
  order to be insensitive to it.  Due to the Reeh-Schlieder-Theorem,
  the second requirement is incompatible with strict locality
  \cite{reeh/schlieder:1961} (cf.~also
  \cite[Theorem~5.3.2]{haag:1996}). The appropriate localization
  concept for detectors turns out to be \emph{almost locality} (or
  \emph{quasilocality of infinite order} \cite{araki/haag:1967})
  meaning that their dislocalized part (outside a finite region of
  radius $r$) falls off more rapidly than any power of $r$. Within
  this setting, Araki and Haag arrived at a decomposition of
  scattering states in terms of energy-momentum eigenstates. They got
  to the following asymptotic relation which holds true for quasilocal
  operators satisfying $C \Omega = C^* \Omega = 0$, $\Omega$ the
  vacuum state, and for arbitrary vectors $\Phi$ and certain specific
  vectors $\Psi$ representing outgoing particle configurations
  \cite[Theorem~4]{araki/haag:1967}:
  \begin{equation}
    \label{eq:araki/haag-result}
    \lim_{t \rightarrow \infty} \bscpx{\Phi}{t^3 C ( h , t )}{\Psi} =
    \sideset{}{'} \sum_{i , j} \int d^3 p \; \Gamma_{i j} ( \pib ) \,
    \bscpx{\Phi}{a_{\! j}^{\dagger \text{out}} ( \pib ) \,
    a_i^{\text{out}} ( \pib )}{\Psi} \, h ( \vib_i ) \text{.}
  \end{equation}
  Here $\Gamma_{i j} ( \pib ) \doteq 8 \pi^3 \bscpx{\pib j}{C ( 0
  )}{\pib \, i}$, and $h$ denotes a function of the velocity that is
  defined by $\vib_i \doteq ( \pib^2 + m_i^2 )^{-1/2} \pib$. The
  indices $i$ and $j$ in the above formula denote the particle types,
  characterized by intrinsic quantum numbers including spin, and
  summation runs over pairs of particles with equal mass: $m_i = m_j$.
  The structure of the right-hand side of this equation is based on
  the \emph{a priori} knowledge of the particle content of the massive
  theory.
  
  Our final goal is to develop a result similar to
  \eqref{eq:araki/haag-result} in a model-independent framework
  without excluding massless states. To this end a considerably
  smaller class of operators is treated as representatives of particle
  detectors. It is based on the observation that, to produce a signal,
  a minimal energy is needed which depends on the characteristics of
  the detector. Henceforth, the relevant operators are required to
  annihilate all physical states with bounded energy below a specific
  threshold.  Again, on account of the Reeh-Schlieder-Theorem, this
  feature is incompatible with strict locality. The construction of
  the algebra of counters thus starts with operators transferring
  energy-momenta appropriate to enforce insensitivity below a certain
  energy bound. Among these the subset of almost local ones is
  selected to localize the states. Upon complementing these
  requirements by an assumption of smoothness with respect to
  Poincar\'e transformations, the actual definition of detectors
  includes the possibility to perform measurements after localization.
  Passing to the limit of asymptotic times in investigating physical
  states of bounded energy by means of these particle counters, one
  arrives at linear functionals $\sigma$ on the algebra of detectors
  which are continuous with respect to suitable topologies and exhibit
  properties of singly localized systems to be interpreted as
  particles. They give rise to certain specific sesquilinear forms on
  the space of localizing operators, the \emph{particle weights}.
  Having thus established a result corresponding to the left-hand side
  of \eqref{eq:araki/haag-result}, we are faced with the problem of
  decomposing the asymptotic functional $\sigma$ in terms of
  energy-momentum eigenstates as suggested by the right-hand side. In
  the corresponding reformulation of \eqref{eq:araki/haag-result} all
  expressions occurring in the integrand apart from $\Gamma_{i j}$ are
  absorbed into measures $\mu_{i , j}$, so that the asymptotic
  functional $\sigma$ is represented as a mixture of linear forms on
  the algebra of detectors with Dirac kets representing improper
  momentum eigenstates:
  \begin{equation}
    \label{eq:araki/haag-reformulated}
    \sigma ( C ) = \sideset{}{'} \sum_{i , j} \int d \mu_{i , j} (
    \pib ) \; \bscpx{\pib j}{C ( 0 )}{\pib \, i} \text{.}
  \end{equation}
  A result of this kind can indeed be established in terms of
  the corresponding sesquilinear forms (particle weights). Here the
  assumption of smoothness with respect to space-time translations
  imposed on the localizing operators turns out to be vital to render
  the concept of particle weights stable in the course of
  disintegration.
  
  The structure of this article is as follows:
  Section~\ref{sec:localizing-operators} develops the concept of
  detectors and investigates suitable topologies. A basic ingredient
  is the interplay between locality and the spectrum condition. In
  Section~\ref{sec:particle-weights} the resulting continuous
  functionals in the dual space are analyzed. Then, on physical
  grounds, a certain subclass is distinguished, arising as asymptotic
  limits of functionals constructed from physical states of bounded
  energy. These limits are to be interpreted as representing
  asymptotic particle configurations. A characteristic of the limiting
  procedure is its ability to directly reproduce charged systems, in
  contrast to the LSZ-theory where charge-carrying unobservable
  operators are necessary. The representations induced by these
  asymptotic functionals are highly reducible. Their disintegration in
  terms of irreducible representations corresponding to \emph{pure}
  particle weights will be presented in a second paper, where a
  formula analogous to equation \eqref{eq:araki/haag-reformulated}
  will be a central result. To facilitate reading, proofs have been
  deferred to Sections \ref{sec:locop-proofs} and
  \ref{sec:weights-proofs}. The Conclusions
  (Section~\ref{sec:conclusions}) put this approach into proper place
  and list topics of further study.

\section{Localizing Operators and Spectral Seminorms}
  \label{sec:localizing-operators}

\subsection{The Algebra of Detectors}
  
  We start this section by giving the exact definitions underlying the
  concept of detectors to be used in the sequel. As mentioned in the
  Introduction, a detector should annihilate states of bounded energy
  below a certain threshold. This feature is implemented by the
  following definition.
  \begin{definition}[Vacuum Annihilation Property]
    \label{def:vacuum-annihilation}
    An operator $A \in \Afrak$ has the \emph{vacuum annihilation
    property}, if, in the sense of operator-valued distributions,
    the mapping
    \begin{equation}
      \label{eq:vacuum-annihilation}
      \Rsone \ni x \mapsto \ax ( A ) \doteq U ( x ) \, A \, U ( x )^*
      \in \Afrak
    \end{equation}
    has a Fourier transform with compact support $\Gamma$ contained in
    the complement of the forward light cone $\fwcone$. The collection
    of all vacuum annihilation operators is a subspace $\Aann$ of
    $\Afrak$.
  \end{definition}
  The support of the Fourier transform of
  \eqref{eq:vacuum-annihilation} is precisely the energy-momen\-tum
  transfer of $A$, and the energy threshold for the annihilation of
  states depends on the distance $d ( \Gamma , \fwcone )$ between
  $\Gamma$ and $\fwcone$ and the position of $\Gamma$.
  
  A second property of detectors should be their localization in
  spacetime. Strict locality being incompatible with the vacuum
  annihilation property, we confine ourselves to almost locality.
  \begin{definition}[Almost Locality]
    \label{def:almost-locality}
    Let $\Oscr_r \doteq \bset{( x^0 , \xib ) \in \Rsone : \abs{x^0} +
    \abs{\xib} < r}$, $r > 0$, denote the double cone (standard
    diamond) with basis $\Oib_r \doteq \bset{\xib \in \Rs : \abs{\xib}
    < r}$. An operator $A \in \Afrak$ is called \emph{almost local},
    if there exists a net $\bset{A_r \in \AOr : r > 0}$ of local
    operators such that
    \begin{equation}
      \label{eq:almost-locality}
      \lim_{r \rightarrow \infty} r^k \norm{A - A_r} = 0
    \end{equation}
    for any $k \in \Nbb_0$. The set of almost local operators is a
    $^*$-subalgebra $\AS$ of $\Afrak$.
  \end{definition}
  The spaces $\Aann$ and $\AS$ are both invariant under Poincar\'e
  transformations $( \Lambda , x ) \in \Poin$; the operator $\aLax ( A
  )$ transfers energy-momentum in $\Lambda \Gamma$ which again
  belongs to the complement of $\fwcone$ like the support $\Gamma$
  pertaining to $A$ itself.
  
  The construction of a subalgebra $\Cfrak$ in $\Afrak$ containing
  those self-adjoint operators to be interpreted as representatives of
  particle detectors is now accomplished in three steps
  (Definitions~\ref{def:annihilator-space}--\ref{def:counter-algebra}),
  where the above characteristics are supplemented by requiring
  smoothness with respect to the Poincar\'e group.
  \begin{definition}
    \label{def:annihilator-space}
    The almost local vacuum annihilation operators $L_0$ with the
    accessory property that the mapping
    \begin{equation}
      \label{eq:operator-differentiability}
      \Poin \ni ( \Lambda , x ) \mapsto \aLax ( L_0 ) \in \Afrak
    \end{equation}
    is infinitely often differentiable with respect to the initial
    topology of the Poincar\'e group $\Poin$ and the norm topology of
    $\Afrak$ constitute a subspace of $\AS \cap \Aann$. The additional
    requirement that all partial derivatives of any order be again
    almost local distinguishes a \emph{vector space} $\Lfrak_0
    \subset \Afrak$.
  \end{definition}
  \begin{remark*}
    \begin{remlist}
    \item Partial derivatives of any order of
      \eqref{eq:operator-differentiability} correspond to vacuum
      annihilation operators inheriting their energy-momentum transfer
      from $L_0$. But they need not necessarily be almost local. By
      virtue of the last condition, the space $\Lfrak_0$ is stable not
      only under Poincar\'e transformations but also under
      differentiation.
    \item A huge number of elements of $\Lfrak_0$ can be constructed
      by regularizing almost local operators with respect to rapidly
      decreasing functions on the Poincar\'e group furnished with the
      Haar measure $\mu$. (Note, that the semi-direct product Lie
      group $\Poin = \Lor \ltimes \Rsone$ is unimodular
      \cite[Proposition~II.29 and Corollary]{nachbin:1965}, since
      $\Lor$ is a simple thus semisimple Lie group
      \cite[Proposition~I.1.6]{helgason:1984}). For $A \in \AS$ the
      operator
      \begin{equation}
        \label{eq:regularized-annihilator}
        A ( F ) \doteq \int d \mu ( \Lambda , x ) \; F ( \Lambda , x )
        \, \aLax ( A )
      \end{equation}
      belongs to $\Lfrak_0$ and transfers energy-momentum in the
      compact set $\Gamma \subseteq \complement \fwcone$, if the
      infinitely often differentiable function $F$ is rapidly
      decreasing on the subgroup $\Rsone$ and compactly supported on
      $\Lor$ (i.\,e., $F \in \Sscr_0 \bigl( \Poin \bigr) = \Sscr_0
            \bigl( \Lor \ltimes \Rsone \bigr)$,
      cf.~\cite{baumgaertel/wollenberg:1992}) and has the additional
      property that the Fourier transforms of the partial functions
      $F_\Lambda (~.~) \doteq F ( \Lambda ,~.~)$ have common support
      in $\Gamma$ for any $\Lambda \in \Lor$.
    \end{remlist}
  \end{remark*}
  The following definition specifies a left ideal $\Lfrak$ of the
  algebra $\Afrak$.
  \begin{definition}
    \label{def:localizer-ideal}
    Let $\Lfrak$ denote the linear span of all operators $L \in
    \Afrak$ of the form $L = A \, L_0$ where $A \in \Afrak$ and $L_0
    \in \Lfrak_0$; i.\,e.,
    \begin{equation*}
      \Lfrak \doteq \Afrak \, \Lfrak_0 = \linhull \bset{A \, L_0 : A
      \in \Afrak , L_0 \in \Lfrak_0} \text{.}
    \end{equation*}
    Then $\Lfrak$ is a left ideal of $\Afrak$, called the \emph{left
    ideal of localizing operators}.
  \end{definition}
  By its very construction, $L \in \Lfrak$ annihilates the vacuum and
  all states of bounded energy below a certain threshold which is
  determined by the minimal threshold associated with a vacuum
  annihilation operator occurring in any of the representations $L =
  \sum_{i = 1}^N A_i \, L_i$, $A_i \in \Afrak$, $L_i \in \Lfrak_0$.
  The elements of $\Lfrak$ are used to define that algebra of
  operators of which the self-adjoint elements are to be interpreted
  as representatives of particle detectors.
  \begin{definition}
    \label{def:counter-algebra}
    Let $\Cfrak$ denote the linear span of all operators $C \in
    \Afrak$ which can be represented in the form $C = {L_1}^* \, L_2$
    with $L_1$, $L_2 \in \Lfrak$; i.\,e.,
    \begin{equation*}
      \Cfrak \doteq \Lfrak^* \, \Lfrak = \linhull \bset{{L_1}^* \, L_2
      : L_1 , L_2 \in \Lfrak} \text{.}
    \end{equation*}
    Then $\Cfrak$ is a $^*$-subalgebra of $\Afrak$, called the
    \emph{algebra of detectors}.
  \end{definition}
  This algebra is smaller than that used by Araki and Haag
  \cite{araki/haag:1967}. It is neither closed in the uniform topology
  of $\Afrak$ nor does it contain a unit. Note that $\Lfrak$ as well
  as $\Cfrak$ are stable under Poincar\'e transformations.

\subsection{Spectral Seminorms on the Algebra of Detectors}
  
  Triggering a detector $C \in \Cfrak$ requires a minimal energy
  $\epsilon$ to be deposited. Take a state $\omega$ of bounded energy
  $E$, then we expect to encounter a finite number of localization
  centers, their number being equal to or less than
  $E/\epsilon$. According to this heuristic picture, placing the
  counter $C$ at every point $\xib$ in space $\Rs$ at given time $t$
  and adding up the corresponding expectation values $\omega \bigl(
  \alpha_{( t , \, \xib )} ( C ) \bigr)$ should result in finiteness
  of the integral 
  \begin{equation}
    \label{eq:heuristic-integral}
    \int_{\Rs} d^s x \; \babs{\omega \bigl( \alpha_{( t , \, \xib )} (
    C ) \bigr)} < \infty \text{.} 
  \end{equation}
  As a matter of fact, the elements of $\Cfrak$ turn out to have this
  property as can be shown by using results of Buchholz
  \cite{buchholz:1990}. The essential ingredient here is the interplay
  between spatial localization and energy bounds, hinting at the
  importance of phase-space properties of quantum field theory.
  \begin{proposition}
    \label{pro:counter-integrals}
    Suppose that $\Delta \subseteq \Rsone$ is a bounded Borel set.
    \begin{proplist}
    \item Let $L \in \Lfrak$ be arbitrary, then the net $\Bset{\ED
      \int_\Kib d^s x \; \aibx ( L^* L ) \; \ED : \Kib \subseteq \Rs
      \text{compact}}$ of operator-valued Bochner integrals
      converges $\sigma$-strongly for $\Kib \nearrow \Rs$, its limit
      being the $\sigma$-weak integral $\int_{\Rs} d^s x \; \ED \aibx
      ( L^* L ) \ED$.
    \item Let $C$ be an arbitrary detector in $\Cfrak$, then the net
      of operator-valued Bochner integrals $\Bset{\ED \int_\Kib d^s x
      \; \aibx ( C ) \; \ED : \Kib \subseteq \Rs \text{compact}}$ is
      $\sigma$-strongly convergent in the limit $\Kib \nearrow \Rs$,
      and its limit is the $\sigma$-weak integral $\int_{\Rs} d^s x \;
      \ED \aibx ( C ) \ED$. Furthermore,
      \begin{equation}
        \label{eq:supremum-seminorm}
        \sup \Bset{\int_{\Rs} d^s x \; \babs{\phi \bigl( \ED \aibx ( C
        ) \ED \bigr)} : \phi \in \BH_{*,1}} < \infty \text{.}
      \end{equation}
    \end{proplist}
  \end{proposition}
  Relation \eqref{eq:supremum-seminorm} constitutes the sharpened
  version of \eqref{eq:heuristic-integral} which was based on
  heuristic considerations.
  
  The preceding result suggests the introduction of topologies on the
  left ideal $\Lfrak$ and on the $^*$-algebra $\Cfrak$, respectively,
  using specific seminorms indexed by bounded Borel subsets $\Delta$
  of $\Rsone$. By their very construction, these are especially
  well-adapted to the problems at hand.
  \begin{subequations}
    \begin{definition}
      \label{def:seminorms}
      \begin{deflist}
      \item The left ideal $\Lfrak$ is equipped with a family of
        seminorms $\qD$ via
        \begin{equation}
          \label{eq:q-seminorm}
          \qDx{L} \doteq \Bnorm{\int_{\Rs} d^s x \; \ED \aibx ( L^* L
          ) \ED}^{1/2} \text{,} \quad L \in \Lfrak \text{.}
        \end{equation}
      \item The $^*$-algebra $\Cfrak$ is furnished with seminorms
        $\pD$ by assigning
        \begin{equation}
          \label{eq:p-seminorm}
          \pDx{C} \doteq \sup \Bset{\int_{\Rs} d^s x \; \babs{\phi
          \bigl( \ED \aibx( C ) \ED \bigr)} : \phi \in \BH_{*,1}}
          \text{,} \quad C \in \Cfrak \text{.}
        \end{equation}
      \item These families of seminorms give $\Lfrak$ and $\Cfrak$ the
        structure of locally convex Hausdorff spaces denoted $( \Lfrak
        , \Tfrak_q )$ and $( \Cfrak , \Tfrak_p )$, respectively.
      \end{deflist}
    \end{definition}

    An alternative expression for $\qD$ can be given in terms of the
    positive cone $\BH^+_*$ in the pre-dual $\BH_*$:
    \begin{equation}
      \label{eq:alternative-q-seminorm}
      \qDx{L}^2 = \sup \Bset{\int_{\Rs} d^s x \; \omega \bigl( \ED
      \aibx ( L^* L ) \ED \bigr) : \omega \in \BH^+_{*,1}} \text{.}
    \end{equation}
  \end{subequations}
  The seminorm properties of $\qD$ and $\pD$ are easily checked. That
  the corresponding families separate elements of $\Lfrak$ and
  $\Cfrak$, respectively, is established as follows: From the very
  definition of the seminorms $\qD$ and $\pD$ we infer that the
  conditions $\qDx{L} = 0$ and $\pDx{C} = 0$, $L \in \Lfrak$, $C \in
  \Cfrak$, by rendering the integrands in
  \eqref{eq:alternative-q-seminorm} and \eqref{eq:p-seminorm}
  identically zero, imply $L \ED = 0$ and $\ED C \ED = 0$ for any
  bounded Borel set $\Delta$, since $\BH^+_{*,1}$ as well as
  $\BH_{*,1}$ are separating sets of functionals for $\BH$. Now,
  states of bounded energy constitute a dense subspace of $\Hscr$ so
  that in conclusion $L = 0$ and $C = 0$.

\subsection{Characteristics of the Spectral Seminorms}
  
  The subsequent investigations very much depend on special properties
  of these seminorms. Here we present only the most important ones,
  deferring their proof as well as the formulation of a couple of
  important Lemmas to Section~\ref{sec:locop-proofs}. Furthermore, we
  are, in the present context, not aiming at utmost generality of
  statements. A more elaborate discussion can be found in
  \cite{porrmann:2000}. $\Delta$, with or without sub- or
  superscripts, generically denotes bounded subsets of the
  energy-momentum space $\Rsone$. The first result concerns the net
  structure of the family of seminorms.
  \begin{proposition}
    \label{pro:seminorm-nets}
    The families of seminorms $\qD$ and $\pD$ on $\Lfrak$ and
    $\Cfrak$, respectively, constitute nets with respect to the
    inclusion relation. For any $\Delta$ and $\Delta'$ we have
    \begin{align*}
      \Delta \subseteq \Delta' \qquad & \Rightarrow \qquad \qDx{L}
      \leqslant \qDprimex{L} \text{,} \quad L \in \Lfrak \text{,} \\
      \Delta \subseteq \Delta' \qquad & \Rightarrow \qquad \pDx{C}
      \leqslant \pDprimex{C} \text{,} \quad C \in \Cfrak \text{.}
    \end{align*}
  \end{proposition}
  Of particular importance in connection with the characteristics of
  particle weights are the invariance of the seminorms with respect to
  spacetime translations as well as special properties of continuity
  and differentiability with respect to the Poincar\'e group.
  \begin{proposition}
    \label{pro:seminorm-translation-invariance}
    Let $x \in \Rsone$ be arbitrary, then
    \begin{proplist}
    \item \hfill $\bqDx{\ax ( L )} = \qDx{L}$ \text{,} \quad $L \in
      \Lfrak$ \text{;}
      \hfill \text{~}
    \item \hfill $\bpDx{\ax ( C )} = \pDx{C}$ \text{,} \quad $C \in
      \Cfrak$ \text{.} \hfill \text{~}
    \end{proplist}
  \end{proposition}
  It is assumed that the automorphism group of Poincar\'e
  transformations acts strongly continuous on the $C^*$-algebra
  $\Afrak$, meaning that the mapping $( \Lambda , x ) \mapsto \aLax (
  A )$ for given $A \in \Afrak$ is continuous with respect to the
  initial topology of $\Poin$ and the uniform topology of $\Afrak$.
  This characteristic is preserved in passing to the subspaces
  $\Lfrak$ and $\Cfrak$ with their respective locally convex
  topologies. For operators in $\Lfrak_0$ even infinite
  differentiability is seen to hold with respect to $( \Lfrak_0 ,
  \Tfrak_q )$. The seminorm topologies not being finer than the norm
  topology, this result is not a corollary of strong continuity of the
  automorphism group.
  \begin{proposition}
    \label{pro:lcs-continuity-differentiability}
    \begin{proplist}
    \item Given $L \in \Lfrak$ and $C \in \Cfrak$, the mappings
      \begin{align*}
        \Poin \ni ( \Lambda , x ) & \mapsto \aLax ( L ) \in \Lfrak
        \text{,} \\
        \Poin \ni ( \Lambda , x ) & \mapsto \aLax ( C ) \in \Cfrak
      \end{align*}
      are continuous in the locally convex spaces $( \Lfrak , \Tfrak_q
      )$ and $( \Cfrak , \Tfrak_p )$.
    \item Given $L_0 \in \Lfrak_0$, the mapping
      \begin{equation*}
        \Poin \ni ( \Lambda , x ) \mapsto \aLax ( L_0 ) \in \Lfrak_0
      \end{equation*}
      is infinitely often differentiable in the locally convex space
      $( \Lfrak_0 , \Tfrak_q )$.  Furthermore, its partial derivatives
      coincide with those arising from the presupposed
      differentiability of this mapping with respect to the uniform
      topology (cf.~Definition~\ref{def:annihilator-space}).
    \end{proplist}
  \end{proposition}

\section{Particle Weights as Asymptotic Plane Waves}
  \label{sec:particle-weights}
  
  We now turn to the investigation of the topological dual space of $(
  \Cfrak , \Tfrak_p)$.
  \begin{definition}
    \label{def:varsigma-continuity}
    \begin{deflist}
    \item The linear functionals on $\Cfrak$, which are continuous
      with respect to the seminorm $\pD$, constitute a normed vector
      space $\CDstar$ via
      \begin{equation*}
        \norm{\varsigma}_\Delta \doteq \sup \bset{\abs{\varsigma ( C
        )} : C \in \Cfrak, \pDx{C} \leqslant 1} \text{,} \quad
        \varsigma \in \CDstar \text{.}
      \end{equation*}
    \item The topological dual of the locally convex space $( \Cfrak ,
      \Tfrak_p)$ is denoted $\Cstar$.
    \end{deflist}
  \end{definition}
  Due to the net property of the family of seminorms $\pD$
  (Proposition~\ref{pro:seminorm-nets}), a linear functional on
  $\Cfrak$ belongs to the topological dual $\Cstar$ if and only if it
  is continuous with respect to a specific one of these seminorms
  \cite[Proposition~1.2.8]{kadison/ringrose:1983}. Hence $\Cstar$ is
  the union of all the spaces $\CDstar$.

\subsection{General Properties}
  \label{subsec:gen-prop}
  
  Before proceeding to extract certain elements from $\Cstar$ to be
  interpreted as representing asymptotic mixtures of particle-like
  entities, we are first going to collect a number of properties
  common to \emph{all} functionals from the topological dual of
  $\Cfrak$. First of all, continuity as established in
  Proposition~\ref{pro:lcs-continuity-differentiability} directly
  carries over.
  \begin{proposition}
    \label{pro:varsigma-continuity}
    Continuous linear functionals $\varsigma \in \Cstar$ have the
    following properties.
    \begin{proplist}
    \item The mapping $( \Lambda , x ) \mapsto \varsigma \bigl(
      {L_1}^* \aLax ( L_2 ) \bigr)$ is continuous for given $L_1$,
      $L_2 \in \Lfrak$.
    \item The mapping $( \Lambda , x ) \mapsto \varsigma \bigl( \aLax
      ( C ) \bigr)$ is continuous for given $C \in \Cfrak$.
    \end{proplist}
  \end{proposition}
  Every \emph{positive} functional $\varsigma$ on the $^*$-algebra
  $\Cfrak = \Lfrak^* \, \Lfrak$ defines a non-negative sesquilinear
  form on $\Lfrak$ through
  \begin{subequations}
    \label{eq:varsigma-quotient-constructions}
    \begin{equation}
      \label{eq:varsigma-sesquilinear-form}
      \scp{~.~}{~.~}_\varsigma : \Lfrak \times \Lfrak \rightarrow \Cbb
      \quad ( L_1 , L_2) \mapsto \scp{L_1}{L_2}_\varsigma \doteq 
      \varsigma ( {L_1}^* L_2 ) \text{,} 
    \end{equation}
    and thus induces a seminorm $\qs$ on $\Lfrak$ via
    \begin{equation}
      \label{eq:varsigma-seminorm}
      \qs : \Lfrak \rightarrow \Rbb_+ \quad L \mapsto \qsx{L} \doteq
      \scp{L}{L}_\varsigma^{1/2} \text{,} 
    \end{equation}
  \end{subequations}
  respectively a norm $\norm{~.~}_\varsigma$ on the quotient space
  $\Lfrak / \Nfrak_\varsigma$, where $\Nfrak_\varsigma$ denotes the
  null space of $\qs$. Using square brackets to designate the cosets
  in $\Lfrak / \Nfrak_\varsigma$, we immediately get the following
  result on differentiability since, by the supposed continuity of
  $\varsigma$, the seminorm $\qs$ is continuous with respect to at
  least one of the seminorms $\qD$.
  \begin{proposition}
    \label{pro:varsigma-differentiability}
    Let $\varsigma$ be a continuous positive functional on the
    $^*$-algebra $\Cfrak$, in short $\varsigma \in \Cstarplus$. Then,
    given $L_0 \in \Lfrak_0$, the mapping
    \begin{equation*}
      \Poin \ni ( \Lambda , x ) \mapsto \bigl[ \aLax ( L_0 )
      \bigr]_\varsigma \in ( \Lfrak_0 / \Nfrak_\varsigma ,
      \norm{~.~}_\varsigma )
    \end{equation*}
    is infinitely often differentiable with respect to the norm
    $\norm{~.~}_\varsigma$.
  \end{proposition}

  The following Cluster Property of positive functionals in $\Cstar$
  is important in characterizing particle weights as being singly
  localized whereas the Spectral Property restricts their possible
  energy-momenta.
  \begin{proposition}[Cluster Property]
    \label{pro:cluster}
    Let $L_i$ and $L'_i$ be elements of $\Lfrak_0$ and let $A_i \in
    \Afrak$, $i = 1$, $2$, be almost local operators, then the
    function
    \begin{equation}
      \label{eq:cluster-function}
      \Rs \ni \xib \mapsto \varsigma \bigl( ( {L_1}^* A_1 L'_1 ) \aibx
      ( {L_2}^* A_2 L'_2 ) \bigr) \in \Cbb
    \end{equation}
    is an element of $L^1 \bigl( \Rs , d^s x \bigr)$ for any
    $\varsigma \in \Cstarplus$ and satisfies
    \begin{equation}
      \label{eq:cluster}
      \int_{\Rs} d^s x \; \babs{\varsigma \bigl( ( {L_1}^* A_1 L'_1 )
      \aibx ( {L_2}^* A_2 L'_2 ) \bigr)} \leqslant
      \norm{\varsigma}_\Delta \, M_\Delta
    \end{equation}
    for any bounded Borel set $\Delta$ for which $\varsigma$ belongs
    to $\CDstar$; $M_\Delta$ is a constant depending on $\Delta$ and
    the operators involved.
  \end{proposition}
  The Spectral Property of functionals $\varsigma \in \Cstar$
  expressed in the subsequent proposition will prove to be of
  importance in defining the energy-momentum of particle weights.
  \begin{proposition}[Spectral Property]
    \label{pro:spectral-property}
    Let $L_1$, $L_2 \in \Lfrak$ and $\varsigma \in \Cstar$. Then
    the support of the Fourier transform of the distribution
    \begin{equation*}
      \Rsone \ni x \mapsto \varsigma \bigl( {L_1}^* \ax ( L_2 ) \bigr)
      \in \Cbb
    \end{equation*}
    is contained in a shifted light cone $\fwcone - q$ for some $q \in
    \fwcone$. More specifically, $q$ is determined by the condition
    $\Delta \subseteq q - \fwcone$, where $\varsigma \in \CDstar$.
  \end{proposition}

\subsection{Asymptotic Functionals}
  \label{subsec:asymptotic_functionals}
  
  Now we turn to functionals in $\Cstar$ arising as temporal limits of
  physical states with bounded energy. The development of such a state
  $\omega \in \Sscr ( \Delta )$, $\Delta$ a bounded Borel set, can be
  explored by considering the following integral
  \begin{equation}
    \label{eq:heuristic-velocity-integral}
    \int_{\Rs} d^s v \; h ( \vib ) \, \omega \bigl( \alpha_{( \tau ,
      \tau \vib )} ( C ) \bigr) \text{,}
  \end{equation}
  where $h$ denotes a bounded measurable function on the unit ball of
  $\Rs$, $\vib$ representing velocity. Apart from $h$,
  \eqref{eq:heuristic-velocity-integral} coincides with the integral
  \eqref{eq:heuristic-integral} encountered in the heuristic
  considerations of Section~\ref{sec:localizing-operators}. The
  investigations of that place
  (cf.~Proposition~\ref{pro:counter-integrals}) imply that
  \eqref{eq:heuristic-velocity-integral} takes on a finite value for
  any counter $C \in \Cfrak$ at given time $\tau$.
  
  The physical interpretation is as follows: Consider a function $h$
  of bounded support $\Vsf \subseteq \Rs \setminus \set{0}$ in
  velocity space, then the integral
  \eqref{eq:heuristic-velocity-integral} corresponds to summing up the
  expectation values of measurements of $C$ in the state $\omega$ at
  time $\tau$, their locus comprising the bounded section $\tau \cdot
  \Vsf$ of configuration space. The distance of this portion from the
  origin together with its total extension increases with time. More
  exactly, the measurements take place in a cone with apex at $0$, its
  orientation being determined by the support of $h$. For different
  times $\tau$ the counter $C$ is set up in specific parts of that
  cone, the extension of which grows as $\abs{\tau}^{s}$ (compensating
  for the quantum mechanical spreading of wave packets), while their
  distance from the origin increases proportionally to $\abs{\tau}$.
  If in the limit of large (positive or negative) times the physical
  state $\omega$ has evolved into a configuration containing a
  particle (incoming or outgoing) travelling with velocity $\vib_0 \in
  \Vsf$, then for a counter $C_0$ sensitive for that specific
  particle the above experimental setup is expected to asymptotically
  yield a constant signal. Mathematically, this corresponds to the
  existence of limits of the above integral at asymptotic times,
  evaluated for the counter $C_0$ and a function $h_0$ with support
  around $\vib_0$.  Thus, the problem has to be settled in which
  (topological) sense such limits can be established. To tackle this
  assignment we turn to a slightly modified version of
  \eqref{eq:heuristic-velocity-integral} involving a certain time
  average for technical reasons.
  \begin{definition}
    \label{def:rho-definition}
    Let $\Delta$ be a bounded Borel subset of\/ $\Rsone$, let $\omega
    \in \Sscr ( \Delta )$ denote a physical state of bounded energy
    and let $\vib \mapsto h ( \vib )$ be a bounded measurable function
    on the unit ball of\/ $\Rs$. Furthermore suppose that $t \mapsto T
    ( t )$ is a continuous real-valued function, approaching $+
    \infty$ or $- \infty$ for asymptotic positive or negative times,
    respectively, not as fast as $\abs{t}$. Then we define a net
    $\bset{ \rho_{h , t} : t \in \Rbb}$ of linear functionals on
    $\Cfrak$ by
    \begin{equation}
      \label{eq:finite-times-functionals}
      \rho_{h , t} ( C ) \doteq T ( t )^{-1} \int_t^{t + T ( t )} d
      \tau \int_{\Rs} d^s x \; h ( \tau^{-1} \xib ) \, \omega \bigl(
      \alpha_{( \tau , \xib )} ( C ) \bigr) \text{,} \quad C \in
      \Cfrak \text{.}
    \end{equation}
  \end{definition}

  The operators $U ( \tau )$ implementing time translations commute
  with $\ED$, a fact that allows \eqref{eq:finite-times-functionals}
  to be re-written for the physical state $\omega (~.~)= \omega \bigl(
  \ED~.~\ED \bigr)$ as
  \begin{equation}
    \label{eq:rho-alt-form}
    \rho_{h , t} ( C ) = T ( t )^{-1} \int_t^{t + T ( t )} d \tau
    \int_{\Rs} d^s x \; h ( \tau^{-1} \xib ) \, \omega \bigl( U ( \tau
    ) \ED \aibx ( C ) \ED U ( \tau )^* \bigr) \text{.}
  \end{equation}
  Since all the functionals $\omega \bigl( U ( \tau )~.~U ( \tau )^*
  \bigr)$, $\tau \in \Rbb$, belong to $\BH_{*,1}$ the left-hand side
  of \eqref{eq:rho-alt-form} can be estimated by
  \begin{multline}
    \label{eq:rho-estimate}
    \babs{\rho_{h , t} ( C )} \leqslant \sup_{\tau \in I_t}
    \Babs{\int_{\Rs} d^s x \; h ( \tau^{-1} \xib ) \, \omega \bigl( U
    ( \tau ) \ED \aibx ( C ) \ED U ( \tau )^* \bigr)} \\
    \leqslant \norm{h}_\infty \sup_{\phi \in \BH_{*,1}} \int_{\Rs} d^s
    x \babs{\phi \bigl( \ED \aibx ( C ) \ED \bigr)} = \norm{h}_\infty
    \pDx{C} \text{,}
  \end{multline}
  where $I_t$ denotes the interval of $\tau$-integration at time $t$.
  This inequality implies that all the functionals $\rho_{h , t}$ are
  continuous with respect to $\pD$, i.\,e., $\rho_{h , t} \in
  \CDstar$. Moreover, \eqref{eq:rho-estimate} is uniform in $t$ so
  that the net $\bset{\rho_{h , t} : t \in \Rbb}$ even turns out to be
  an equicontinuous subset of $\Cstar$. The Theorem of
  Alao\u{g}lu-Bourbaki \cite[Theorem~8.5.2]{jarchow:1981} then tells
  us that this net is relatively compact with respect to the weak
  topology, leading to the following fundamental result.
  \begin{theorem}[Existence of Limits]
    \label{the:singular-limits}
    Under the assumptions of Definition~\ref{def:rho-definition} the
    net $\bset{\rho_{h , t} : t \in \Rbb} \subseteq \CDstar$ possesses
    weak limit points in $\Cstar$ at asymptotic times. This means that
    there exist functionals $\sigma_{h , \omega}^{( + )}$ and
    $\sigma_{h , \omega}^{( - )}$ on $\Cfrak$ together with
    corresponding subnets $\bset{\rho_{h , t_\iota} : \iota \in J}$
    and $\bset{\rho_{h , t_\kappa} : \kappa \in K}$, where $\lim_\iota
    t_\iota = + \infty$ and $\lim_\kappa t_\kappa = - \infty$, such
    that for arbitrary $C \in \Cfrak$
    \begin{subequations}
      \label{eq:singular-limits}
      \begin{align}
        \rho_{h , t_\iota} ( C ) & \underset{\iota}{\longrightarrow}
        \sigma_{h , \omega}^{( + )} ( C ) \text{,} \\ 
        \rho_{h , t_\kappa} ( C ) & \underset{\kappa}{\longrightarrow}
        \sigma_{h , \omega}^{( - )} ( C ) \text{.}
      \end{align}
    \end{subequations}
  \end{theorem}

  The heuristic picture laid open above suggests that in theories with
  a complete particle interpretation the net of functionals actually
  converges, but the proof is still lacking. One has to assure that in
  the limit of large times multiple scattering does no longer withhold
  the measurement results $\rho_{h , t} ( C )$ from growing stable.
  Another problem is the possible vanishing of the limit functionals
  on all of $\Cfrak$, a phenomenon we anticipate to encounter in
  theories without a particle interpretation (e.\,g.~the generalized
  free field). They both seem to be connected with the problem of
  asymptotic completeness of quantum field theoretic models and
  reasonable criteria to restrict their phase space structure. The
  denomination of asymptotic functionals by $\sigma$ reflects their
  \emph{singular} nature: the values that the functionals $\rho_{h ,
  t}$ return for finite times $t$ when applied to the identity
  operator $\unit$ (which is not contained in $\Cfrak$) are divergent
  as $\abs{t}^s$ at asymptotic times.
  
  The convergence problem as yet only partially solved in the sense of
  Theorem~\ref{the:singular-limits}, one can nevertheless establish a
  number of distinctive properties of the limit functionals. An
  immediate first consequence of the above construction is the
  following proposition.
  \begin{proposition}[Positivity and Continuity of Limits]
    \label{pro:positivity-of-limits}
    Suppose that $\Delta$ is a bounded Borel subset of\/ $\Rsone$,
    $\omega \in \Sscr ( \Delta )$ a physical state of bounded energy
    and $h \in L^\infty ( \Rs , d^s x )$ a non-negative function. Then
    the limit functionals $\sigma$ of the net $\bset{\rho_{h , t} : t
    \in \Rbb}$ are positive elements of $\CDstar$:
    \begin{subequations}
      \begin{align}
        \babs{\sigma ( C )} & \leqslant \norm{h}_\infty \, \pDx{C}
        \text{,} \quad C \in \Cfrak \text{;} \\
        0 & \leqslant \sigma ( C ) \text{,} \quad C \in \Cfrak^+
        \text{.}
      \end{align}
    \end{subequations}
  \end{proposition}
  The proofs of Propositions~\ref{pro:translation-invariance} and
  \ref{pro:lower-bounds} require a special asymptotic behaviour of the
  function $h \in L^\infty ( \Rs , d^s v )$: it has to be continuous,
  approximating a constant value in the limit $\abs{\vib} \rightarrow
  \infty$, i.\,e., $h - M_h \in C_0 ( \Rs )$ for a suitable constant
  $M_h$; these functions constitute a subspace of $C ( \Rs )$ that
  will be denoted $C_{0 , c} ( \Rs )$.
  \begin{proposition}[Translation Invariance]
    \label{pro:translation-invariance}
    Let $\Delta \subseteq \Rsone$ be a bounded Borel set, let $\omega
    \in \Sscr ( \Delta )$ and $h \in C_{0 , c} ( \Rs )$. Then the
    limit functionals $\sigma$ of the net$\bset{\rho_{h , t} : t \in
    \Rbb}$ are invariant under spacetime translations, i.\,e., $
    \sigma \bigl( \ax ( C ) \bigr) = \sigma ( C )$ for any $C \in
    \Cfrak$ and any $x \in \Rsone$.
  \end{proposition}
  The last property of those special elements $\sigma \in \CDstarplus$
  arising as limits of nets of functionals $\bset{\rho_{h , t_\iota} :
  \iota \in J}$ complements the Cluster Property~\ref{pro:cluster}.
  It asserts, given certain specific operators $C \in \Cfrak$, the
  existence of a \emph{lower} bound for the integral of $\xib \mapsto
  \sigma \bigl( C^* \aibx ( C ) \bigr)$.
  \begin{proposition}[Existence of Lower Bounds]
    \label{pro:lower-bounds}
    Let $C \in \Cfrak$ be a counter which has the property that the
    function $\xib \mapsto \bpDx{C^* \aibx ( C )}$ is integrable
    (cf.~Lemma~\ref{lem:varsigmanet-convergence}). Let furthermore
    $\sigma \in \CDstarplus$ be the limit of a net of functionals
    $\bset{\rho_{h , t_\iota} : \iota \in J}$, where the velocity
    function $h$ is non-negative and belongs to $C_{0 , c} ( \Rs )$.
    Then
    \begin{equation}
      \label{eq:lower-bounds}
      \babs{\sigma ( C )}^2 \leqslant \norm{h}_\infty \int_{\Rs} d^s x
      \; \sigma \bigl( C^* \aibx ( C ) \bigr) \text{.}
    \end{equation}
  \end{proposition}

\subsection{Particle Weights}
  \label{subsec:particle-weights}
    
  The features of limit functionals collected thus far suggest their
  interpretation as representing mixtures of particle-like quantities
  with sharp energy-momentum: being translationally invariant
  according to Proposition~\ref{pro:translation-invariance}, they
  appear as plane waves, i.\,e., energy-momentum eigenstates; on the
  other hand, they are singly localized at all times by
  Proposition~\ref{pro:cluster}, a feature to be expected for a
  particle-like system. In addition,
  Proposition~\ref{pro:spectral-property} determines the
  energy-momentum spectrum. Systems of this kind shall be summarized
  under the concept of \emph{particle weights}, a term reflecting the
  connection to the notion of \emph{weights} or \emph{extended
    positive functionals} in the theory of $C^*$-algebras. Their
  definition goes back to Dixmier \cite[Section~I.4.2]{dixmier:1981}
  (cf.~also \cite[Section~5.1]{pedersen:1979} and
  \cite{pedersen:1966}) and designates functionals on the positive
  cone $\Afrak^+$ of a $C^*$-algebra $\Afrak$ that can attain infinite
  values, a property they share with the singular functionals of
  Theorem~\ref{the:singular-limits}. Their domain $\Cfrak$ does not
  comprise the unit $\unit$ of the quasi-local algebra, since the
  defining approximation yields the value $\sigma ( \unit ) = +
  \infty$.
    
  As mentioned on page~\pageref{eq:varsigma-quotient-constructions},
  every positive functional $\sigma$ on $\Cfrak = \Lfrak^* \, \Lfrak$
  defines a non-negative sesquilinear form $\scp{~.~}{~.~}_\sigma$ on
  $\Lfrak \times \Lfrak$ that induces a seminorm $q_\sigma$ on
  $\Lfrak$ and a norm $\norm{~.~}_\sigma$ on the corresponding
  quotient $\Lfrak / \Nfrak_\sigma$ with respect to the null space
  $\Nfrak_\sigma$ of $q_\sigma$. Taking advantage of these
  constructions, we shall depart from functionals and proceed to
  sesquilinear forms, which are intimately connected with
  representations of $\Afrak$. The following definition reformulates
  the results on limit functionals within this setting.
  \begin{definition}
    \label{def:particle-weight}
    A particle weight is a non-trivial, non-negative sesquilinear form
    on $\Lfrak$, denoted $\scp{~.~}{~.~}$, which induces a seminorm
    $q_w$ with null space $\Nfrak_w$ on the ideal $\Lfrak$ and a norm
    $\norm{~.~}_w$ on the quotient $\Lfrak / \Nfrak_w$ complying with
    the following conditions:
    \begin{proplist}
    \item for any $L_1$, $L_2 \in \Lfrak$ and $A \in \Afrak$ one has
      $\scp{L_1}{A \, L_2} = \scp{A^* L_1}{L_2}$;
    \item for given $L \in \Lfrak$ the following mapping is continuous
      with respect to $\qw$:
      \begin{equation*}
        \Poin \ni ( \Lambda , x ) \mapsto \aLax ( L ) \in \Lfrak
        \text{;}
      \end{equation*}
    \item given $L_0 \in \Lfrak_0$, the mapping
      \begin{equation*}
        \Poin \ni ( \Lambda , x ) \mapsto \bigl[ \aLax ( L_0 )
        \bigr]_w \in ( \Lfrak_0 / \Nfrak_w , \norm{~.~}_w )
      \end{equation*}
      is infinitely often differentiable with respect to the norm
      $\norm{~.~}_w$
      (cf.~Proposition~\ref{pro:varsigma-differentiability}).
    \item $\scp{~.~}{~.~}$ is invariant with respect to spacetime
      translations $x \in \Rsone$, i.\,e.,
      \begin{equation*}
        \bscp{\ax ( L_1 )}{\ax ( L_2 )} = \scp{L_1}{L_2} \text{,}
        \quad L_1 \text{, } L_2 \in \Lfrak \text{,}
      \end{equation*}
      and the $(s + 1)$-dimensional Fourier transforms of the
      distributions $x \mapsto \bscp{L_1}{\ax ( L_2 )}$ have support
      in a shifted forward light cone $\fwcone - q$ with $q \in
      \fwcone$.
    \end{proplist}
  \end{definition}
  Note that we did not use any of the $\qD$-topologies on $\Lfrak$ in
  formulating this definition, for properties given in terms of these
  will in general get lost in the disintegration to be expounded in a
  second article. Instead, the relevant topologies are those induced
  on $\Lfrak$ by the sesquilinear form itself. The constituent
  properties are preserved under the operations of addition and of
  multiplication by positive numbers. Therefore, the totality of
  particle weights, supplemented by the trivial form, proves to be a
  positive (proper convex) cone, denoted $\Wecm$, in the linear space
  of all sesquilinear forms on $\Lfrak$
  (cf.~\cite{peressini:1967,asimow/ellis:1980}). It is by no means
  self-evident how to pass from a sesquilinear form of the above type
  to a positive linear functional on $\Cfrak$. It requires additional
  restrictive assumptions on the structure of $\Cfrak$ to render the
  definition of the associated functional unambiguous.
    
  A completely equivalent characterization of particle weights can be
  given in terms of representations $( \pi_w , \Hscr_w )$ of the
  quasi-local algebra $\Afrak$, obtained by a GNS-construction
  (cf.~\cite[Theorem~3.2]{pedersen:1966} and
  \cite[Proposition~5.1.3]{pedersen:1979}).
  \begin{theorem}
    \label{the:particle-weight}
    \begin{theolist}
    \item To any particle weight $\scp{~.~}{~.~}$ there corresponds a
      non-zero, non-degenerate representation $( \pi_w , \Hscr_w )$ of
      the quasi-local $C^*$-algebra $\Afrak$ with the following
      properties:
      \begin{theosublist}
      \item there exists a linear mapping $\ket{~.~}$ from $\Lfrak$
        onto a dense subspace of $\Hscr_w$:
        \begin{equation*}
          \ket{~.~} : \Lfrak \rightarrow \Hscr_w \qquad L \mapsto
          \ket{L} \text{,}
        \end{equation*}
        such that the representation $\pi_w$ is given by
        \begin{equation*}
          \pi_w ( A ) \ket{L} = \ket{A L} \text{,} \quad A \in
          \Afrak \text{,} \quad L \in \Lfrak \text{;}
        \end{equation*}
      \item the following mapping is continuous for given $L \in
        \Lfrak$:
        \begin{equation}
          \label{eq:ket-continuous}
          \Poin \ni ( \Lambda , x ) \mapsto \bket{\aLax ( L )} \in
          \Hscr_w \text{;}
        \end{equation}
      \item for $L_0 \in \Lfrak_0$ the mapping
        \eqref{eq:ket-continuous} is infinitely often
        differentiable;
      \item there exists a strongly continuous unitary representation
        $x \mapsto U_w ( x )$ of spacetime translations $x \in \Rsone$
        on $\Hscr_w$ defined by
        \begin{equation*}
          U_w ( x ) \ket{L} \doteq \bket{\ax ( L )} \text{,} \quad L
          \in \Lfrak \text{,}
        \end{equation*}
        with spectrum in a displaced forward light cone $\fwcone - q$,
        $q \in \fwcone$.
      \end{theosublist}
    \item Any representation $( \pi_w , \Hscr_w )$ which has the above
      characteristics defines a particle weight through the scalar
      product on $\Hscr_w$.
    \end{theolist}
  \end{theorem}
  \begin{remark*}
    The unitaries $U_w ( x )$ implement the group $\bset{\ax : x \in
    \Rsone}$ of automorphisms of $\Afrak$ in $( \Hscr_w , \pi_w )$
    via
    \begin{equation} 
      \label{eq:auto-implement}
      U_w ( x ) \pi_w ( A ) {U_w ( x )}^* = \pi_w \bigl( \ax ( A )
      \bigr) \text{,} \quad A \in \Afrak \text{,} \quad x \in \Rsone
      \text{.}
    \end{equation}
  \end{remark*}

  The energy-momentum transfer of operators in $\Lfrak$ determines the
  spectral subspace to which the corresponding vectors in $\Hscr_w$
  pertain. Moreover, particle weights enjoy a Cluster Property
  parallel to that established in Proposition~\ref{pro:cluster} for
  positive functionals in $\CDstar$.
  \begin{proposition}[Spectral Subspaces]
    \label{pro:spectral-subspace}
    Let $L \in \Lfrak$ have energy-momentum transfer in the Borel
    subset $\Delta'$ of\/ $\Rsone$. Then, in the representation $(
    \pi_w , \Hscr_w )$ corresponding to the particle weight
    $\scp{~.~}{~.~}$, the vector $\ket{L}$ belongs to the spectral
    subspace pertaining to $\Delta'$ with respect to the generator of
    the intrinsic unitary representation $x \mapsto U_w ( x )$ of
    spacetime translations:
    \begin{equation}
      \label{eq:energy-momentum-subspace}
      \ket{L} = \EwDprime \ket{L} \text{.}
    \end{equation}
  \end{proposition}
  \begin{proposition}[Cluster Property for Particle Weights]
    \label{pro:weights-cluster}
    Let $L_i$ and $L'_i$ be elements of $\Lfrak_0$ with
    energy-momentum transfer in $\Gamma_i$ respectively $\Gamma'_i$,
    and let $A_i \in \Afrak$, $i = 1$, $2$, be almost local.
    Then, for the particle weight $\scp{~.~}{~.~}$ with
    GNS-representation $( \pi_w , \Hscr_w )$,
    \begin{equation*}
      \Rs \ni \xib \mapsto \bscp{{L_1}^* A_1 L'_1}{\aibx ( {L_2}^* A_2
      L'_2 )} = \bscpx{{L_1}^* A_1 L'_1}{U_w ( \xib )}{{L_2}^* A_2
      L'_2} \in \Cbb
    \end{equation*}
    is a function in $L^1 \bigl( \Rs , d^s x \bigr)$.
  \end{proposition}

  At this point a brief comment on the notation chosen seems
  appropriate (cf.~\cite{buchholz/porrmann/stein:1991}). We
  deliberately utilize the typographical token $\ket{~.~}$ introduced
  by Dirac \label{page:dirac-kets} \cite[\S\,23]{dirac:1958} for ket
  vectors describing improper momentum eigenstates $\ket{\pib}$, $\pib
  \in \Rs$. These act as distributions on the space of momentum wave
  functions with values in the physical Hilbert space $\Hscr$, thereby
  presupposing a superposition principle to hold without limitations.
  This assumption collapses in an infraparticle situation as described
  in the Introduction. In contrast to this, the \emph{pure} particle
  weights, that will appear in connection with elementary physical
  systems, are seen to be associated with sharp momentum and yet
  capable of describing infraparticles. Here the operators $L \in
  \Lfrak$ take on the role of the previously mentioned momentum space
  wave functions in that they localize the particle weight in order to
  produce a normalizable vector $\ket{L}$ in the Hilbert space
  $\Hscr_w$. This in turn substantiates the terminology introduced in
  Definition~\ref{def:localizer-ideal}.  Describing elementary
  physical systems, pure particle weights should give rise to
  irreducible representations of the quasi-local algebra, which
  motivates the subsequent definition. It is supplemented by a notion
  of boundedness which is in particular shared by functionals $\sigma$
  arising from the construction expounded in
  Subsection~\ref{subsec:asymptotic_functionals}
  (cf.~Lemma~\ref{lem:Delta-bound}).
  \begin{definition}
    \label{def:weight-classification}
    A particle weight is said to be
    \begin{deflist}
    \item \emph{pure}, if the corresponding representation $( \pi_w ,
      \Hscr_w )$ is irreducible;
    \item \emph{$\Delta$-bounded}, if to any bounded Borel subset
      $\Delta'$ of $\Rsone$ there exists another such set $\Deltabar
      \supseteq \Delta + \Delta'$ so that the GNS-representation $(
      \pi_w , \Hscr_w )$ of the particle weight and the defining
      representation are connected by the following inequality,
      valid for any $A \in \Afrak$,
      \begin{equation}
        \label{eq:Delta-boundedness}
        \norm{\EwDprime \pi_w ( A ) \EwDprime} \leqslant c \cdot
        \norm{\EDbar A \EDbar}
      \end{equation}
      with a suitable positive constant $c$ that is independent of the
      Borel sets involved. Obviously, $\Delta$ ought to be a bounded
      Borel set as well.
    \end{deflist}
  \end{definition}
  \begin{lemma}
    \label{lem:Delta-bound}
    Any positive asymptotic functional $\sigma \in \CDstar$,
    constructed according to Theorem~\ref{the:singular-limits} under
    the assumptions of Proposition~\ref{pro:positivity-of-limits},
    gives rise to a $\Delta$-bounded particle weight
    $\scp{~.~}{~.~}_\sigma$.
  \end{lemma}
  The above concepts are important in connection with the
  disintegration theory (pure particle weights) and with the local
  normality of representations ($\Delta$-boundedness). Both will be
  the topic of the forthcoming second article.

\section{Proofs for Section~\ref{sec:localizing-operators}}
  \label{sec:locop-proofs}

  In a first step to prove the validity of
  \eqref{eq:heuristic-integral} for detectors in $\Cfrak$, we consider
  detectors built from special operators $L_0 \in \Lfrak_0$ having
  energy-momentum transfer in \emph{convex} sets.
  \begin{proposition}
    \label{pro:harmonic-analysis}
    Let $E (~.~)$ be the spectral resolution corresponding to the
    unitary representation $x \mapsto U ( x )$ of spacetime
    translations and let $L_0 \in \Lfrak_0$ have energy-momentum
    transfer in the compact and convex subset $\Gamma$ of $\complement
    \fwcone$. Then the net of operator-valued Bochner integrals
    $\Bset{\ED \int_\Kib d^s x \; \aibx ( {L_0}^* L_0 ) \; \ED : \Kib
    \subseteq \Rs \text{compact}}$ is $\sigma$-strongly convergent
    in $\BH$ as $\Kib \nearrow \Rs$ for any bounded Borel set $\Delta
    \subseteq \Rsone$. Its limit is the $\sigma$-weak integral of the
    $\BH^+$-valued mapping $\xib \mapsto \ED \aibx ( {L_0}^* L_0 )
    \ED$ and satisfies the following estimate for suitable $N ( \Delta
    , \Gamma ) \in \Nbb$ depending on $\Delta$ and $\Gamma$:
    \begin{equation}
      \label{eq:commutator-integral-estimate}
      \Bnorm{\int_{\Rs} d^s x \; \ED \aibx ( {L_0}^* L_0 ) \ED}
      \leqslant N ( \Delta , \Gamma ) \int_{\Rs} d^s x \;
      \bnorm{\bcomm{\aibx ( L_0 )}{{L_0}^*}} \text{.}
    \end{equation}
  \end{proposition}
  \begin{remark*}
    The integral on the right-hand side of
    \eqref{eq:commutator-integral-estimate} is finite due to almost
    locality of $L_0$: Let $A$ and $B$ be almost local with
    approximating nets $\bset{A_r \in \AOr : r > 0}$ and $\bset{B_r
    \in \AOr : r > 0}$, respectively. Since $\Oscr_r$ and $\Oscr_r +
    2 \xib$ are spacelike separated for $r \leqslant \abs{\xib}$,
    $\xib \in \Rs \setminus \set{\zeroib}$, one has
    \begin{equation}
      \label{eq:commutator-norm-estimate}
      \bnorm{\bcomm{\aibtwox ( A )}{B}} \leqslant 2 \, \bigl(
      \norm{A - A_{\abs{\xib}}} \, \norm{B} + \norm{A -
      A_{\abs{\xib}}} \, \norm{B - B_{\abs{\xib}}} + \norm{A} \,
      \norm{B - B_{\abs{\xib}}} \bigr) \text{,}
    \end{equation}
    implying integrability of the left-hand side over all of $\Rs$.
  \end{remark*}
  \begin{proof}
    $\Delta$ being a bounded Borel set with closure
    $\overline{\Delta}$, there exists, due to compactness and
    convexity of $\Gamma$, a number $n \in \Nbb$ such that $(
    \overline{\Delta} + \Gamma_n ) \cap \fwcone = \emptyset$ is
    satisfied, where $\Gamma_n$ denotes the $n$-fold sum of the
    compactum $\Gamma$. The spectrum condition then entails $E (
    \overline{\Delta} + \Gamma_n ) = 0$. As a consequence, for
    arbitrary $\xib_1$, \dots, $\xib_n \in \Rs$, the product $\prod_{i
    = 1}^n \alpha_{\xib_i} ( L_0 )$ annihilates any state with
    energy-momentum in $\Delta$ so that $\ED$ turns out to be a
    projection with range in the intersection of the kernels of
    $n$-fold products of the above kind. An application of
    \cite[Lemma~2.2]{buchholz:1990} then yields the estimate
    \begin{equation}
      \label{eq:harmonic-analysis-commutator-estimate}
      \Bnorm{\ED \int_\Kib d^s x \; \aibx ( {L_0}^* L_0 ) \; \ED}
      \leqslant ( n - 1 ) \, \int_{\Rs} d^s x \; \bnorm{\bcomm{\aibx (
      L_0 )}{{L_0}^*}}
    \end{equation}
    for any compact $\Kib$, where the right-hand side is finite due to
    almost locality of $L_0$. The positive operators on the left-hand
    side thus constitute an increasing net that is bounded and
    converges $\sigma$-strongly to its least upper bound in $\BH^+$
    \cite[Lemma~2.4.19]{bratteli/robinson:1987}.
  
    The $\sigma$-weak topology on $\BH$ is induced by the positive
    normal functionals $\psi$ in the cone $\BH^+_*$. Thus,
    integrability of $\xib \mapsto \ED \aibx ( {L_0}^* L_0 ) \ED$ with
    respect to the $\sigma$-weak topology is implied by integrability
    of all of the corresponding functions $\xib \mapsto \babs{\psi
    \bigl( \ED \aibx ( {L_0}^* L_0 ) \ED \bigr)} = \psi \bigl( \ED
    \aibx ( {L_0}^* L_0 ) \ED \bigr)$. Given any compact subset $\Kib$
    of $\Rs$, we have, by
    \eqref{eq:harmonic-analysis-commutator-estimate},
    \begin{multline*}
      \int_\Kib d^s x \; \babs{\psi \bigl( \ED \aibx ( {L_0}^* L_0 )
      \ED \bigr)} = \psi \Bigl( \int_\Kib d^s x \; \ED \aibx ( {L_0}^*
      L_0 ) \ED \Bigr) \\
      \leqslant \norm{\psi} \Bnorm{\int_\Kib d^s x \; \ED \aibx (
      {L_0}^* L_0 ) \ED} \leqslant \norm{\psi} \; ( n - 1 ) \, \int_{\Rs}
      d^s x \; \bnorm{\bcomm{\aibx ( L_0 )}{{L_0}^*}}
    \end{multline*}
    so that, as a consequence of the Monotone Convergence Theorem
    \cite[II.2.7]{fell/doran:1988a}, the functions $\xib \mapsto
    \babs{\psi \bigl( \ED \aibx ( {L_0}^* L_0 ) \ED \bigr)}$ indeed
    turn out to be integrable. The integral of $\xib \mapsto \ED \aibx
    ( {L_0}^* L_0 ) \ED$ over $\Rs$ with respect to the $\sigma$-weak
    topology thus exists \cite[II.6.2]{fell/doran:1988a} and, by
    application of Lebesgue's Dominated Convergence Theorem
    \cite[II.5.6]{fell/doran:1988a}, is seen to be the $\sigma$-weak
    limit of the net of compactly supported integrals. It thus
    coincides with the $\sigma$-strong limit established above and,
    according to \eqref{eq:harmonic-analysis-commutator-estimate},
    satisfies \eqref{eq:commutator-integral-estimate} with $N ( \Delta
    , \Gamma ) \doteq n - 1$.
  \end{proof}
  The special result of Proposition~\ref{pro:harmonic-analysis} for
  $L_0 \in \Lfrak_0$ can now easily be generalized to arbitrary $L \in
  \Lfrak$ and $C \in \Cfrak$ as laid down in
  Proposition~\ref{pro:counter-integrals}.
  \begin{proof}[Proposition~\ref{pro:counter-integrals}]
    \begin{prooflist}
    \item By partition of unity applied to the Fourier transform of
      the mapping $x \mapsto U ( x ) \, L_0 \, U ( x )^*$, any $L_0
      \in \Lfrak_0$ is seen to be representable as a finite sum of
      operators $L_i \in \Lfrak_0$ transferring energy-momentum in
      compact and \emph{convex} subsets $\Gamma_i$ of $\complement
      \fwcone$.  Accordingly, any $L \in \Lfrak$ can be written as $L
      = \sum_{j = 1}^m A_j L_j$ with $A_j \in \Afrak$ and $L_j \in
      \Lfrak_0$ of this special kind. The relation
      \begin{equation*}
        L^* L \leqslant 2^{m - 1} \Bigl( \sup_{1 \leqslant j \leqslant
        m} \norm{A_j}^2 \Bigr) \sum_{j = 1}^m {L_j}^* L_j \text{,}
      \end{equation*}
      implies for any compact $\Kib \subseteq \Rs$
      \begin{equation*}
        \Bnorm{\ED \negthinspace \int_\Kib \negthickspace d^s x \;
        \aibx ( L^* L ) \, \ED} \leqslant 2^{m - 1} \Bigl( \sup_{1
        \leqslant j \leqslant m} \norm{A_j}^2 \Bigr) \sum_{j = 1}^m
        \Bnorm{\ED \negthinspace \int_\Kib \negthickspace d^s x \;
        \aibx ( {L_j}^* L_j ) \, \ED} \text{,}
      \end{equation*}
      and the statement now follows by use of the arguments given in
      the proof of Proposition~\ref{pro:harmonic-analysis} which
      apply to the vacuum annihilation operators $L_j$.
    \item By polarization, an arbitrary element $C_0 = {L_1}^* L_2 \in
      \Cfrak$ with $L_1$, $L_2 \in \Lfrak$ can be written as a linear
      combination of four positive operators of the form $L^* L$, $L
      \in \Lfrak$. So the problem of $\sigma$-strong convergence of
      the net of Bochner integrals constructed with $C_0$ reduces to
      the case already established in the first part. Polar
      decomposition of the normal functional $\phi$
      \cite[Theorem~III.4.2(i), Proposition~III.4.6]{takesaki:1979},
      yields a partial isometry $V \in \BH$ and a \emph{positive}
      normal functional $\abs{\phi}$ subject to $\norm{\abs{\phi}} =
      \norm{\phi}$ such that $\phi(~.~) = \abs{\phi} (~.~V )$. This
      entails for arbitrary $\xib \in \Rs$ and $\lambda > 0$ the
      estimate
      \begin{multline*}
        2 \, \babs{\phi \bigl( \ED \aibx ( C_0 ) \ED \bigr)} = 2 \,
        \babs{\abs{\phi} \bigl( \ED \aibx ( {L_1}^* L_2 ) \ED V
        \bigr)} \\
        \mspace{-45mu} \leqslant 2 \, \sqrt{\abs{\phi} \bigl( \ED
        \aibx ( {L_1}^* L_1 ) \ED \bigr)} \sqrt{\abs{\phi} \bigl( V^*
        \ED \aibx ( {L_2}^* L_2 ) \ED V \bigr)} \\
        \leqslant \lambda^{-1} \abs{\phi} \bigl( \ED \aibx ( {L_1}^*
        L_1 ) \ED \bigr) + \lambda \, \abs{\phi} \bigl( V^* \ED \aibx
        ( {L_2}^* L_2 ) \ED V \bigr) \text{.}
      \end{multline*}
      Both sides of this inequality are integrable over $\Rs$, since
      from part (i) we infer
      \begin{multline*}
        2 \int_{\Rs} d^s x \; \babs{\phi \bigl(\ED \aibx ( C_0 ) \ED
        \bigr)} \\
        \leqslant \lambda^{-1} \norm{\phi} \; \Bnorm{\int_{\Rs} d^s x
        \; \ED \aibx ( {L_1}^* L_1 ) \ED} + \lambda \, \norm{\phi} \;
        \Bnorm{\int_{\Rs}d^s x \; \ED \aibx ( {L_2}^* L_2 ) \ED}
        \text{,}
      \end{multline*}
      by noting that normal functionals and $\sigma$-weak integrals
      commute due to \cite[Proposition~II.5.7 adapted to integrals in
      locally convex spaces]{fell/doran:1988a}. Taking the infimum
      with respect to $\lambda$, one arrives at
      \begin{multline}
        \label{eq:C-integrability}
        \int_{\Rs} d^s x \; \babs{\phi \bigl( \ED \aibx ( C_0 ) \ED
        \bigr)} \\
        \leqslant \norm{\phi} \; \Bnorm{\int_{\Rs} d^s x \; \ED \aibx (
        {L_1}^* L_1 ) \ED}^{1/2} \; \Bnorm{\int_{\Rs}d^s x \; \ED \aibx (
        {L_2}^* L_2 ) \ED}^{1/2} \text{.}
      \end{multline}
      This relation being valid for any normal functional $\phi \in
      \BH_*$, the asserted existence of the $\sigma$-weak integral is
      established for $C_0 \in \Cfrak$, which obviously coincides with
      the $\sigma$-strong limit of the corresponding net. As a
      by-product of \eqref{eq:C-integrability} we have
      \begin{multline}
        \label{eq:supremum-estimate-prepared}
        \sup \Bset{\int_{\Rs} d^s x \; \babs{\phi \bigl( \ED \aibx (
        C_0 ) \ED \bigr)} : \phi \in \BH_{*,1}} \\
        \leqslant \Bnorm{\int_{\Rs} d^s x \; \ED \aibx ( {L_1}^* L_1 )
        \ED}^{1/2} \; \Bnorm{\int_{\Rs}d^s x \; \ED \aibx ( {L_2}^*
        L_2 ) \ED}^{1/2} \text{.}
      \end{multline}
      These results are readily extended to establish part (ii) in
      the general case, since any $C \in \Cfrak$ is a linear
      combination of operators of the form of $C_0$.
    \end{prooflist}
    \renewcommand{\qed}{}
  \end{proof}

  What follows are proofs as well as statements of seminorm
  properties.
  \begin{proof}[Proposition~\ref{pro:seminorm-nets}]
    Let $L \in \Lfrak$ and let $\Delta \subseteq \Delta'$, then
    \begin{equation*}
      0 \leqslant \int_{\Rs} d^s x \; \ED \aibx ( L^* L ) \ED
      \leqslant \int_{\Rs} d^s x \; \EDprime \aibx ( L^* L ) \EDprime
      \text{,}
    \end{equation*}
    which by \eqref{eq:q-seminorm} implies $\qDx{L}^2 \leqslant
    \qDprimex{L}^2$. In case of the $\pD$-topologies, note that for
    any Borel set $\Delta$ the functional $\phi^{\ED}$, defined
    through $\phi^{\ED} (~.~) \doteq \phi (\ED~.~\ED)$, belongs to
    $\BH_{*,1}$ if $\phi$ does. From this we infer, since moreover
    $\ED = \ED \EDprime = \EDprime \ED$ is implied by $\Delta
    \subseteq \Delta'$, that
    \begin{multline*}
      \Bset{\int_{\Rs} d^s x \; \babs{\phi \bigl( \ED \aibx( C ) \ED
          \bigr)} : \phi \in \BH_{*,1}} \\ 
      \subseteq \Bset{\int_{\Rs} d^s x \; \babs{\phi \bigl( \EDprime
          \aibx( C ) \EDprime \bigr)} : \phi \in \BH_{*,1}}  
    \end{multline*}
    for any $C \in \Cfrak$ and thus, by \eqref{eq:p-seminorm}, that
    $\pDx{C} \leqslant \pDprimex{C}$.
  \end{proof}
  The statements of the following Lemmas can easily be established by
  use of the integral representations for the seminorms.
  \begin{lemma}
    \label{lem:delta-annihilation}
    \begin{proplist}
    \item For $L \in \Lfrak$ with $L \ED = 0$, one has $\qDx{L} = 0$.
    \item If \/$C \in \Cfrak$ satisfies $\ED C \ED = 0$, then $\pDx{C} =
      0$.
    \end{proplist}
  \end{lemma}
  \begin{lemma}
    \label{lem:basic-estimate}
    \begin{proplist}
    \item For any $L \in \Lfrak$ and $A \in \Afrak$ 
      \begin{equation}
        \label{eq:ideal-estimate}
        \qDx{A L} \leqslant \norm{A} \, \qDx{L} \text{.}
      \end{equation}
    \item Let $L_i$, $i = 1$, $2$, be operators in $\Lfrak$
      having energy-momentum transfer in $\Gamma_i \subseteq \Rsone$,
      and let $\Delta_i$ denote Borel subsets of\/ $\Rsone$ containing
      $\Delta + \Gamma_i$, respectively. Then, for any $A \in \Afrak$,
      \begin{equation}
        \label{eq:counter-estimate}
        \pDx{{L_1}^* A L_2} \leqslant \norm{E ( \Delta_1 ) A E (
        \Delta_2 )} \, \qDx{L_1} \, \qDx{L_2} \text{.} 
      \end{equation}
    \end{proplist}
  \end{lemma}
  \begin{proof}
    \begin{prooflist}
    \item For $L \in \Lfrak$ and $A \in \Afrak$ the relation $L^* A^*
      A L \leqslant \norm{A}^2 L^* L$ implies
      \begin{equation*}
        \int_{\Rs} d^s x \; \omega \bigl( \ED \aibx ( L^* A^* A L ) \ED
        \bigr) \leqslant \norm{A}^2 \int_{\Rs} d^s x \; \omega \bigl(
        \ED \aibx ( L^* L ) \ED \bigr)
      \end{equation*}
      for any $\omega \in \BH^+_{*,1}$ and thus, by virtue of
      \eqref{eq:alternative-q-seminorm},
      \begin{multline*}
        \qDx{A L}^2 = \sup \Bset{\int_{\Rs} d^s x \; \omega \bigl( \ED
        \aibx ( L^* A^* A L ) \ED \bigr) : \omega \in \BH^+_{*,1}} \\
        \leqslant \norm{A}^2 \sup \Bset{\int_{\Rs} d^s x \; \omega
        \bigl( \ED \aibx ( L^* L ) \ED \bigr) : \omega \in
        \BH^+_{*,1}} = \norm{A}^2 \qDx{L}^2 \text{.}
      \end{multline*}
    \item Let $\phi$ be a normal functional on $\BH$ with $\norm{\phi}
      \leqslant 1$. By polar decomposition, there exist a partial
      isometry $V$ and a positive normal functional $\abs{\phi}$ with
      $\norm{\abs{\phi}} \leqslant 1$ such that $\phi (~.~) =
      \abs{\phi} (~.~V )$. Then for any $\xib \in \Rs$
      \begin{multline*}
        \babs{\phi \bigl( \ED \aibx ( {L_1}^* A L_2 ) \ED \bigr)} =
        \abs{\phi} \bigl( \ED \aibx ( {L_1}^* ) E ( \Delta_1 ) \aibx (
        A ) E ( \Delta_2 ) \aibx ( L_2 ) \ED V \bigr) \\
        \leqslant \negthinspace \norm{E ( \Delta_1 ) \aibx ( A ) E (
        \Delta_2 )} \negthinspace \sqrt{\abs{\phi} \bigl( \ED \aibx (
        {L_1}^* L_1 ) \ED \bigr)} \negthinspace \sqrt{\abs{\phi}
        \bigl( V^* \ED \aibx ( {L_2}^* L_2 ) \ED V \bigr)} \text{,}
      \end{multline*}
      and the method used in the proof of
      Proposition~\ref{pro:counter-integrals} yields,
      in analogy to \eqref{eq:supremum-estimate-prepared},
      \begin{multline*}
        \pDx{{L_1}^* A L_2} = \sup \Bset{\int_{\Rs} d^s x \;
        \babs{\phi \bigl( \ED \aibx ( {L_1}^* A L_2 ) \ED \bigr)} :
        \phi \in \BH_{*,1}} \\
        \leqslant \norm{E ( \Delta_1 ) A E ( \Delta_2 )} \, \qDx{L_1}
        \, \qDx{L_2} \text{,}
      \end{multline*}
      where use was made of \eqref{eq:p-seminorm}.
    \end{prooflist}
    \renewcommand{\qed}{}
  \end{proof}
  An immediate consequence of the second part of this Lemma is
  \begin{corollary}
    \label{cor:sesquilinear-product}
    The sesquilinear mapping on the topological product of the locally
    convex space $( \Lfrak , \Tfrak_q )$ with itself, defined by 
    \begin{equation*}
      \Lfrak \times \Lfrak \ni ( L_1 , L_2 ) \mapsto {L_1}^* L_2 \in
      \Cfrak \text{,}
    \end{equation*}
    is continuous with regard to the respective locally convex
    topologies.
  \end{corollary}
  \begin{lemma}
    \label{lem:involution-invariance}
    \begin{proplist}
    \item For any operator $L \in \Lfrak$ there holds the relation
      \begin{equation*}
        \pDx{L^* L} = \qDx{L}^2 \text{.}
      \end{equation*}
    \item Let $C$ be an element of $\Cfrak$, then
      \begin{equation*}
        \pDx{C^*} = \pDx{C} \text{.}
      \end{equation*}
    \end{proplist}
  \end{lemma}
  \begin{proof}
    \begin{prooflist}
    \item By \eqref{eq:alternative-q-seminorm} and
      \eqref{eq:p-seminorm}, we have for $L \in \Lfrak$
      \begin{multline*}
        \qDx{L}^2 = \sup \Bset{\int_{\Rs} d^s x \; \omega \bigl( \ED
        \aibx ( L^* L ) \ED \bigr) : \omega \in \BH^+_{*,1}} \\
        \leqslant \sup \Bset{\int_{\Rs} d^s x \; \babs{\phi \bigl( \ED
        \aibx ( L^* L ) \ED \bigr)} : \phi \in \BH_{*,1}} = \pDx{L^*
        L} \text{,}
      \end{multline*}
      whereas the reverse inequality is a consequence of
      Lemma~\ref{lem:basic-estimate}.
    \item Note that $\BH_{*,1}$ is invariant under the operation of
      taking adjoints defined by $\psi \mapsto \psi^*$ with $\psi^* (
      A ) \doteq \overline{\psi ( A^* )}$, $A \in \BH$, for any linear
      functional $\psi$ on $\BH$. Thus
      \begin{multline*}
        \pDx{C^*} = \sup \Bset{\int_{\Rs} d^s x \; \babs{\phi
        \bigl( \ED \aibx ( C^* ) \ED \bigr)} : \phi \in \BH_{*,1}} \\
        = \sup \Bset{\int_{\Rs} d^s x \; \babs{\phi^* \bigl( \ED \aibx
        ( C ) \ED \bigr)} : \phi \in \BH_{*,1}} = \pDx{C}  
      \end{multline*}
      for any $C \in \Cfrak$.
    \end{prooflist}
    \renewcommand{\qed}{}
  \end{proof}
  The above results are used in the proofs of
  Propositions~\ref{pro:seminorm-translation-invariance} and
  \ref{pro:lcs-continuity-differentiability}.
  \begin{proof}[Proposition~\ref{pro:seminorm-translation-invariance}]
    $\BH_{*,1}$ as well as its intersection $\BH^+_{*,1}$ with the
    positive cone $\BH^+_*$ are invariant under the mapping $\psi
    \mapsto \psi^U$ defined on $\BH^*$ for any unitary operator $U \in
    \BH$ by $\psi^U (~.~) \doteq \psi ( U~.~U^* )$.
    \begin{prooflist}
    \item Now, $\ax ( L^* L ) = U_t \aibx ( L^* L ) {U_t}^*$ for any
      $x = ( t , \xib ) \in \Rsone$. This implies
      \begin{equation*}
        \omega \bigl( \ED \aiby \bigl( \ax ( L^* L ) \bigr) \ED \bigr)
        = \omega \bigl( U_t \ED \alpha_{\xib + \yib} ( L^* L ) \ED
        {U_t}^* \bigr)
      \end{equation*}
      for any $\yib \in \Rs$ and any $\omega \in \BH_*$. Henceforth
      \begin{multline*}
        \int_{\Rs} d^s y \; \omega \bigl( \ED \aiby \bigl(
        \ax ( L^* L ) \bigr) \ED \bigr) = \int_{\Rs} d^s y \; \omega
        \bigl( U_t \ED \alpha_{\xib + \yib} ( L^* L ) \ED {U_t}^*
        \bigr) \\ 
        = \int_{\Rs} d^s y \; \omega \bigl( U_t \ED \aiby ( L^*
        L ) \ED {U_t}^* \bigr)
      \end{multline*}
      so that, by virtue of \eqref{eq:alternative-q-seminorm}, for any
      $L \in \Lfrak$
      \begin{multline*}
        \bqDx{\ax ( L )}^2 = \sup \Bset{\int_{\Rs} d^s y \; \omega
          \bigl( \ED \aiby \bigl( \ax ( L^* L ) \bigr) \ED \bigr) :
          \omega \in \BH^+_{*,1}} \\
        \mspace{50mu} = \sup \Bset{\int_{\Rs} d^s y \; \omega \bigl( U_t
          \ED \aiby ( L^* L ) \ED {U_t}^* \bigr) : \omega \in
          \BH^+_{*,1}} \\
        = \sup \Bset{\int_{\Rs} d^s y \; \omega \bigl( \ED \aiby (
          L^* L ) \ED \bigr) : \omega \in \BH^+_{*,1}} = \qDx{L}^2
        \text{.}
      \end{multline*}
    \item The same argument applies to $\pD$, so that
      \begin{multline*}
        \bpDx{\ax ( C )} = \sup \Bset{\int_{\Rs} d^s y \; \babs{\phi
        \bigl( \ED \aiby \bigl( \ax ( C ) \bigr) \ED \bigr)} :
        \phi \in \BH_{*,1}} \\
        \mspace{50mu} = \sup \Bset{\int_{\Rs} d^s y \; \babs{\phi
        \bigl( U_t \ED \aiby ( C ) \ED {U_t}^* \bigr)} : \phi
        \in \BH_{*,1}} \\
        = \sup \Bset{\int_{\Rs} d^s y \; \babs{\phi \bigl( \ED
        \aiby ( C ) \ED \bigr)} : \phi \in \BH_{*,1}} = \pDx{C}
      \end{multline*}
    for $C \in \Cfrak$. 
    \end{prooflist}
    \renewcommand{\qed}{}
  \end{proof}
  \begin{proof}%
         [Proposition~\ref{pro:lcs-continuity-differentiability}]
    \begin{prooflist}
    \item Continuity of the mapping $( \Lambda , x ) \mapsto \aLax ( L
      )$ in the locally convex space $( \Lfrak , \Tfrak_q )$ means
      continuity on all of $\Poin$ with respect to each of the
      seminorms $\qD$, and this in turn is established by
      demonstrating continuity in $( \unit , 0 )$ for arbitrary
      localizing operators $L$, since $\Lfrak$ is invariant under
      Poincar\'e transformations.
    
      Let the Borel subset $\Delta$ of $\Rsone$ be arbitrary but fixed
      and consider $L' \in \Lfrak_0$ having energy-momentum transfer
      $\Gamma$ which, under sufficiently small Poincar\'e
      transformations, stays within a compact and convex subset
      $\widehat{\Gamma}$ of $\complement \fwcone$. Then relation
      \eqref{eq:commutator-integral-estimate} in
      Proposition~\ref{pro:harmonic-analysis} tells us that
      \begin{equation}
        \label{eq:applied-harmonic-analysis}
        \bqDx{\aLax ( L' ) - L'}^2 \leqslant N ( \Delta ,
        \widehat{\Gamma} ) \int_{\Rs} d^s y \;
        \bnorm{\bcomm{\aiby \bigl( \aLax ( L' ) - L'
        \bigr)}{\bigl( \aLax ( L' ) - L' \bigr)^*}} \text{.}
      \end{equation}
      Exploiting almost locality of the operators on the right-hand
      side to control the norm of the large radius part of the
      function $\yib \mapsto \bnorm{\bcomm{\aiby \bigl( \aLax (
      L' ) - L' \bigr)}{\bigl( \aLax ( L' ) - L' \bigr)^*}}$, one
      can establish the existence of an integrable majorizing
      function, independent of $( \Lambda , x )$ in a small
      neighbourhood of $( \unit , 0 )$. Moreover, this integrand
      vanishes pointwise in the limit $( \Lambda , x ) \rightarrow (
      \unit , 0 )$ due to strong continuity of the automorphism group.
      Therefore, by Lebesgue's Dominated Convergence Theorem, we infer
      that for any sequence of transformations approaching $( \unit ,
      0 )$
      \begin{equation*}
        \lim_{n \rightarrow \infty} \bqDx{\alpha_{( \Lambda_n , x_n )}
        ( L' ) - L'} = 0 \text{.}
      \end{equation*}
      Since $\Poin$ as a topological space satisfies the first axiom
      of countability, this suffices to establish continuity of the
      mapping $( \Lambda ,x ) \mapsto \aLax ( L' )$ in $( \unit , 0 )$
      with respect to the $\qD$-topology. An arbitrary operator $L \in
      \Lfrak$ can be represented as $L = \sum_{i = 1}^N A_i L'_i$,
      where $L'_i \in \Lfrak_0$ comply with the above assumptions on
      $L'$ and the operators $A_i$ belong to the quasi-local algebra
      $\Afrak$ for any $i=1$, \dots, $N$. According to
      Lemma~\ref{lem:basic-estimate}, one can then derive an estimate
      for $\bqDx{\aLax ( L ) - L}$ in terms of $\bqDx{\aLax ( L'_i ) -
        L'_i}$ and $\bnorm{\aLax ( A_i ) - A_i}$, so that the mapping
      $( \Lambda , x ) \mapsto \aLax ( L )$ is seen to be continuous
      in $( \unit , 0 )$ with respect to $\qD$ for arbitrary $L \in
      \Lfrak$.
    
      Continuity of the mapping $( \Lambda , x ) \mapsto \aLax ( C )$
      in the locally convex space $( \Cfrak , \Tfrak_p )$ is
      equivalent to its being continuous with respect to all seminorms
      $\pD$. This problem again reduces to the one already solved
      above, if one takes into account the shape of general elements
      of $\Cfrak$ according to Definition~\ref{def:counter-algebra} in
      connection with Corollary~\ref{cor:sesquilinear-product}.
    \item Let $L_0 \in \Lfrak_0$ be given and consider the
      parametrization $\hib \mapsto ( \Lambda_\hib , x_\hib )$ of an
      open neighbourhood of $( \Lambda_\zeroib , x_\zeroib )$. With
      respect to this, the derivative at $( \Lambda_\zeroib ,
      x_\zeroib )$ of the mapping $( \Lambda , x ) \mapsto \aLax ( L_0
      )$ yields an approximation that differs from the actual change
      with $\hib$ by the residual term
      \begin{subequations}
        \begin{equation}
          \label{eq:Ldiff-residual}
          R ( \hib ) = \alpha_{( \Lambda_\hib , x_\hib )} ( L_0 ) -
          \alpha_{( \Lambda_\zeroib , x_\zeroib )} ( L_0 ) - \sum_{i ,
          j} h_j \, C_{i j} ( \zeroib ) \, \alpha_{( \Lambda_\zeroib
          , x_\zeroib )} \bigl( \delta^i ( L_0 ) \bigr) \text{.}
        \end{equation}
        Here $\hib \mapsto C_{i j} ( \hib )$ are analytic functions
        and $\delta^i ( L_0 )$ denote the partial derivatives
        associated with $L_0$ at $( \unit , 0) \in \Poin$. Due to the
        presupposed differentiability with respect to the uniform
        topology this residual term satisfies
        \begin{equation}
          \label{eq:Ldiff-residual-limit}
          \lim_{\hib \rightarrow \zeroib} \, \abs{\hib}^{-1} \bnorm{R
          ( \hib )} = 0 \text{,}
        \end{equation}
        a relation which now has to be shown to stay true with the
        norm replaced by any of the seminorms $\qD$. According to the
        Mean Value Theorem in a version generalized to vector-valued
        differentiable mappings on manifolds, for small $\hib$
        \begin{equation*}
          \alpha_{( \Lambda_\hib , x_\hib )} ( L_0 ) - \alpha_{(
          \Lambda_\zeroib , x_\zeroib )} ( L_0 ) = \int_0^1 d
          \vartheta \; \sum_{i , j} h_j \, C_{i j} ( \vartheta \hib )
          \, \alpha_{( \Lambda_{\vartheta \hib} , x_{\vartheta \hib}
          )} \bigl( \delta^i ( L_0 ) \bigr) \text{,}
        \end{equation*}
        where the integral is to be understood with respect to the
        norm topology of $\Afrak$. Thus \eqref{eq:Ldiff-residual} can
        be re-written as
        \begin{equation}
          \label{eq:residual-mean-value}
          R ( \hib ) = \sum_{i , j} h_j \int_0^1 d \vartheta \; \Bigl(
          C_{i j} ( \vartheta \hib ) \, \alpha_{( \Lambda_{\vartheta
          \hib} , x_{\vartheta \hib} )} \bigl( \delta^i ( L_0 ) \bigr)
          - C_{i j} ( \zeroib ) \, \alpha_{( \Lambda_\zeroib ,
          x_\zeroib )} \bigl( \delta^i ( L_0 ) \bigr) \Bigr) \text{.}
        \end{equation}
      \end{subequations}
      As a consequence of the first statement, the
      integrand on the right-hand side is continuous with respect to
      all seminorms $\qD$, so that the integral exists in the
      completion of the locally convex space $( \Lfrak , \Tfrak_q )$.
      By application of \cite[II.6.2 and 5.4]{fell/doran:1988a}, this
      leads to
      \begin{equation*}
        \abs{\hib}^{-1} \bqDx{R ( \hib )} \leqslant \sum_{i , j}
        \max_{0 \leqslant \vartheta \leqslant 1} \BqDx{C_{i j} (
        \vartheta \hib ) \, \alpha_{( \Lambda_{\vartheta \hib} ,
        x_{\vartheta \hib} )} \bigl( \delta^i ( L_0 ) \bigr) - C_{i j}
        ( \zeroib ) \, \alpha_{( \Lambda_\zeroib , x_\zeroib )} \bigl(
        \delta^i ( L_0 ) \bigr)} \text{,}
      \end{equation*}
      where evidently the right-hand side vanishes for $\hib
      \rightarrow \zeroib$. This establishes differentiability of the
      mapping $( \Lambda , x ) \mapsto \aLax ( L_0 )$ for $L_0 \in
      \Lfrak_0$ with respect to $( \Lfrak , \Tfrak_q )$ and, by
      \eqref{eq:Ldiff-residual}, coincidence of the derivatives with
      those presupposed by the definition of $\Lfrak_0$.
    \end{prooflist}
    \renewcommand{\qed}{}
  \end{proof}

  Based on the continuity result of
  Proposition~\ref{pro:lcs-continuity-differentiability}, it is
  possible to construct new elements of (the completions of) the
  locally convex spaces by way of integration.
  \begin{lemma}
    \label{lem:Poin-Bochner-integrals}
    Let the function $F \in L^1 \bigl( \Poin , d \mu ( \Lambda , x )
    \bigr)$ have compact support $\Ssf$.
    \begin{proplist}
    \item To any $L_0 \in \Lfrak_0$ another operator in $\Lfrak_0$ is
      associated by
      \begin{equation}
        \label{eq:L0-integral}
        \alpha_F ( L_0 ) \doteq \int d \mu ( \Lambda , x ) \; F (
        \Lambda , x ) \, \aLax ( L_0 ) \text{.}
      \end{equation}
    \item If $L \in \Lfrak$ and $C \in \Cfrak$, then
      \begin{subequations}
        \label{eq:LC-integral}
        \begin{align}
          \label{eq:L-integral}
          \alpha_F ( L ) & \doteq \int d \mu ( \Lambda , x ) \; F (
          \Lambda , x ) \, \aLax ( L ) \text{,} \\
          \label{eq:C-integral}
          \alpha_F ( C ) & \doteq \int d \mu ( \Lambda , x ) \; F (
          \Lambda , x ) \, \aLax ( C )
        \end{align}
      \end{subequations}
      exist as integrals in the completions of $( \Lfrak , \Tfrak_q )$
      and $( \Cfrak , \Tfrak_p )$, respectively, and satisfy
      \begin{subequations}
        \label{eq:LC-integral-estimate}
        \begin{align}
          \label{eq:L-integral-estimate}
          \bqDx{\alpha_F ( L )} & \leqslant \norm{F}_1 \sup_{( \Lambda
          , x ) \in \Ssf} \bqDx{\aLax ( L )} \text{,} \\
          \label{eq:C-integral-estimate}
          \bpDx{\alpha_F ( C )} & \leqslant \norm{F}_1 \sup_{( \Lambda
          , x ) \in \Ssf} \bpDx{\aLax ( C )} \text{.}
        \end{align}
      \end{subequations}
    \end{proplist}
  \end{lemma}
  \begin{proof}
    \begin{prooflist}
    \item By assumption, $( \Lambda , x ) \mapsto \abs{F ( \Lambda , x
      )} \, \norm{\aLax ( L_0 )} = \abs{F ( \Lambda , x )} \,
      \norm{L_0}$ is an integrable majorizing function for the
      integrand of \eqref{eq:L0-integral}, so $\alpha_F ( L_0 )$
      exists as a Bochner integral in $\Afrak$. The same holds true
      with $L_0$ replaced by approximating local operators $L_{0 , r}
      \in \AOr$. Due to compactness of $\Ssf$, these integrals belong
      to the local algebras $\Afrak \bigl( \Oscr_{r ( \Ssf )} \bigr)$
      pertaining to standard diamonds in $\Rsone$ with an
      $s$-dimensional basis of radius $r ( \Ssf ) \doteq a ( \Ssf ) r
      + b ( \Ssf )$, $a ( \Ssf )$ and $b ( \Ssf )$ suitable
      positive constants. Now,
      \begin{equation*}
        \alpha_F ( L_0 ) - \alpha_F ( L_{0,r} ) = \int_\Ssf d \mu (
        \Lambda , x ) \; F ( \Lambda , x ) \bigl( \aLax ( L_0 ) -
        \aLax ( L_{0,r} ) \bigr)
      \end{equation*}
      can be estimated for any $k \in \Nbb$ ($\mu ( \Ssf )$ is the
      measure of the compact set $\Ssf$) by
      \begin{equation*}
        0 \leqslant r ( \Ssf )^k \bnorm{\alpha_F ( L_0 ) - \alpha_F (
        L_{0,r} )} \leqslant \mu ( \Ssf ) \norm{F}_1 \bigl( a ( \Ssf )
        r + b ( \Ssf ) \bigr)^k \norm{L_0 - L_{0,r}} \text{.}
      \end{equation*}
      Due to almost locality of $L_0$, the right-hand side vanishes in
      the limit of large $r$ so that $\alpha_F ( L_0 )$ itself is
      almost local with approximating net $\Bset{\alpha_F ( L_{0,r} )
      \in \Afrak \bigl( \Oscr_{r ( \Ssf )} \bigr) : r > 0}$.
      
      Let $\Gamma \subseteq \complement \fwcone$ denote the
      energy-momentum transfer of the vacuum annihilation operator
      $L_0$, then, by the Fubini Theorem
      \cite[II.16.3]{fell/doran:1988a}, for any $g \in L^1 \bigl(
      \Rsone , d^{s + 1} y \bigr)$
      \begin{equation*}
        \int_{\Rsone} d^{s + 1} y \; g(y) \, \alpha_y \bigl( \alpha_F
        ( L_0 ) \bigr) = \int_\Ssf d \mu ( \Lambda , x ) \; F (
        \Lambda , x ) \int_{\Rsone} d^{s + 1} y \; g(y) \, \alpha_y
        \bigl( \aLax ( L_0 ) \bigr) \text{.}
      \end{equation*}
      If the support of the Fourier transform $\tilde{g}$ of $g$
      satisfies $\supp \tilde{g} \subseteq \bigcap_{( \Lambda , x )
      \in \Ssf} \complement( \Lambda \Gamma )$, the inner integrals
      on the right-hand side vanish for any $( \Lambda , x ) \in
      \Ssf$ so that
      \begin{equation*}
        \int_{\Rsone} d^{s + 1} y \; g(y) \, \alpha_y \bigl( \alpha_F
        ( L_0 ) \bigr) = 0 \text{.}
      \end{equation*}
      The energy-momentum transfer of $\alpha_F ( L_0 )$ is thus
      contained in $\bigcup_{( \Lambda , x ) \in \Ssf} \Lambda
      \Gamma$, a compact subset of $\complement \fwcone$, and
      $\alpha_F ( L_0 )$ turns out to be a vacuum annihilation
      operator.
      
      Finally, infinite differentiability of the mapping $( \Lambda ,
      x ) \mapsto \aLax \bigl( \alpha_F ( L_0 ) \bigr)$ with respect
      to the uniform topology has to be established. Using the
      notation introduced in the proof of the second part of
      Proposition~\ref{pro:lcs-continuity-differentiability}, we get
      the counterparts of equations \eqref{eq:Ldiff-residual} and
      \eqref{eq:residual-mean-value} with $\aLax ( L_0 )$ (likewise
      infinitely often differentiable) in place of $L_0$:
      \begin{subequations}
        \begin{multline}
          \label{eq:gen-residual-diff-mean-value}
          R^{( \Lambda , x )} ( \hib ) \\
          = \alpha_{( \Lambda_\hib , x_\hib )} \bigl( \aLax ( L_0 )
          \bigr) - \alpha_{( \Lambda_\zeroib , x_\zeroib )} \bigl(
          \aLax ( L_0 ) \bigr) - \sum_{i , j} h_j \, C_{i j} ( \zeroib
          ) \, \alpha_{( \Lambda_\zeroib , x_\zeroib )} \bigl(
          \delta^i \bigl( \aLax ( L_0 ) \bigr) \bigr) \\
          = \sum_{i , j} h_j \int_0^1 \negthickspace d \vartheta \;
          \Bigl( C_{i j} ( \vartheta \hib ) \, \alpha_{(
          \Lambda_{\vartheta \hib} , x_{\vartheta \hib} )} \bigl(
          \delta^i \bigl( \aLax ( L_0 ) \bigr) \bigr) - C_{i j} (
          \zeroib ) \, \alpha_{( \Lambda_\zeroib , x_\zeroib )} \bigl(
          \delta^i \bigl( \aLax ( L_0 ) \bigr) \bigr) \Bigr) \text{.}
          \\
          ~
        \end{multline}
        Upon multiplication by the compactly supported function $F$,
        integration of all parts in this sequence of equations (which
        are continuous in $( \Lambda , x )$) yields, :
        \begin{multline}
          \label{eq:gen-residual-integrated}
          \int d \mu ( \Lambda , x ) \; F ( \Lambda , x ) \, R^{(
          \Lambda , x )} ( \hib ) \\
          \shoveleft{= \alpha_{( \Lambda_\hib , x_\hib )} \bigl(
          \alpha_F ( L_0 ) \bigr) - \alpha_{( \Lambda_\zeroib ,
          x_\zeroib )} \bigl( \alpha_F ( L_0 ) \bigr)} \\
          \shoveright{- \sum_{i , j} h_j \, C_{i j} ( \zeroib ) \int d
          \mu ( \Lambda , x ) \; F ( \Lambda , x ) \, \alpha_{(
          \Lambda_\zeroib , x_\zeroib )} \bigl( \delta^i \bigl( \aLax
          ( L_0 ) \bigr) \bigr)} \\
          \shoveleft{= \sum_{i , j} h_j \int d \mu ( \Lambda , x )
          \int_0^1 d \vartheta \; F ( \Lambda , x ) \, \cdot} \\
          \cdot \Bigl( C_{i j} ( \vartheta \hib ) \,
          \alpha_{( \Lambda_{\vartheta \hib} , x_{\vartheta \hib} )}
          \bigl( \delta^i \bigl( \aLax ( L_0 ) \bigr) \bigr) - C_{i j}
          ( \zeroib ) \, \alpha_{( \Lambda_\zeroib , x_\zeroib )}
          \bigl( \delta^i \bigl( \aLax ( L_0 ) \bigr) \bigr) \Bigr)
          \text{.}
        \end{multline}
      \end{subequations}
      The second equation suggests to consider its last term as the
      derivative at $( \Lambda_\zeroib , x_\zeroib )$ of the mapping
      in question and the left-hand side as the residual term in the
      chosen parametrization. This interpretation is correct if the
      norm of the left-hand side multiplied by $\abs{\hib}^{-1}$
      vanishes in the limit $\hib \rightarrow \zeroib$. This is
      true since, by the third part of
      \eqref{eq:gen-residual-integrated},
      \begin{multline*}
        \abs{\hib}^{-1} \Bnorm{\int d \mu ( \Lambda , x ) \; F (
        \Lambda , x ) \, R^{( \Lambda , x )} ( \hib )} \\
        \shoveleft{\leqslant \sum_{i , j} \int d \mu ( \Lambda , x )
        \int_0^1 d \vartheta \; \abs{F ( \Lambda , x )} \, \cdot} \\
        \cdot \Bnorm{C_{i j} ( \vartheta \hib ) \,
        \alpha_{( \Lambda_{\vartheta \hib} , x_{\vartheta \hib} )}
        \bigl( \delta^i \bigl( \aLax ( L_0 ) \bigr) \bigr) - C_{i j} (
        \zeroib ) \, \alpha_{( \Lambda_\zeroib , x_\zeroib )} \bigl(
        \delta^i \bigl( \aLax ( L_0 ) \bigr) \bigr)} \text{,}
      \end{multline*}
      and the right-hand side is easily seen to tend to $0$ as $\hib
      \rightarrow \zeroib$, by an application of Lebesgue's Dominated
      Convergence Theorem \cite[II.5.6]{fell/doran:1988a} upon noting
      the pointwise vanishing of the integrand in this limit. Now,
      \begin{equation*}
        \delta^i \bigl( \aLax ( L_0 ) \bigr) = \sum_k D_{i k} (
        \Lambda , x ) \aLax \bigl( \delta^k ( L_0 ) \bigr)
      \end{equation*}
      with analytic functions $D_{i k}$. Thus, the derivative of $(
      \Lambda , x ) \mapsto \aLax \bigl( \alpha_F ( L_0 ) \bigr)$ at
      $( \Lambda_\zeroib , x_\zeroib )$ in the $j$-th direction
      resulting from the second part of
      \eqref{eq:gen-residual-integrated} can be written
      \begin{multline*}
        \sum_i C_{i j} ( \zeroib ) \int d \mu ( \Lambda , x ) \; F (
        \Lambda , x ) \, \alpha_{( \Lambda_\zeroib , x_\zeroib )}
        \bigl( \delta^i \bigl( \aLax ( L_0 ) \bigr) \bigr) \\
        = \sum_{i k} C_{i j} ( \zeroib ) D_{i k} ( \Lambda , x )
        \alpha_{( \Lambda_\zeroib , x_\zeroib )} \Bigl( \int d \mu (
        \Lambda , x ) \; F ( \Lambda , x ) \, \aLax \bigl( \delta^k (
        L_0 ) \bigr) \Bigr) \\
        = \sum_{i k} C_{i j} ( \zeroib ) D_{i k} ( \Lambda , x )
        \alpha_{( \Lambda_\zeroib , x_\zeroib )} \bigl( \alpha_F
        \bigl( \delta^k ( L_0 ) \bigr) \bigr) \text{.}
      \end{multline*}
      The operators $\delta^k ( L_0 )$ belong to $\Lfrak_0$.  So, as a
      result of the above reasoning, $\alpha_F \bigl( \delta^k ( L_0 )
      \bigr)$ is an almost local vacuum annihilation operator which in
      addition is differentiable. Thus, derivatives of $( \Lambda , x
      ) \mapsto \aLax \bigl( \alpha_F ( L_0 ) \bigr)$ of arbitrary
      order exist and belong to $\Lfrak_0$.
    \item By Proposition~\ref{pro:lcs-continuity-differentiability},
      the mappings $( \Lambda , x ) \mapsto \aLax ( L )$ and $(
      \Lambda , x ) \mapsto \aLax ( C )$ are continuous with respect
      to the uniform topology and all the $\qD$- and $\pD$-topologies,
      hence bounded on the compact support of $F$. This implies their
      measurability together with the fact that their product with the
      integrable function $F$ is majorized in each case by a multiple
      of $\abs{F}$. As a consequence, the integrals $\alpha_F ( L )$
      and $\alpha_F ( C )$ exist in the completions of the locally
      convex spaces $( \Lfrak , \Tfrak_q )$ and $( \Cfrak , \Tfrak_p
      )$, respectively, and \eqref{eq:LC-integral-estimate} is an
      immediate upshot of \cite[II.6.2 and 5.4]{fell/doran:1988a}.
    \end{prooflist}
    \renewcommand{\qed}{}
  \end{proof}
  There exists a version of the second part of this Lemma for
  Lebesgue-integrable functions on $\Rsone$.
  \begin{lemma}
    \label{lem:Lebesgue-Bochner-integrals}
    Let $L \in \Lfrak$ and let $g \in L^1 \bigl( \Rsone , d^{s + 1} x
    \bigr)$. Then
    \begin{equation}
      \label{eq:alpha-g-def}
      \alpha_g ( L ) \doteq \int_{\Rsone} d^{s + 1} x \; g ( x ) \,
      \ax ( L )
    \end{equation}
    is an operator in the completion of $( \Lfrak , \Tfrak_q )$,
    satisfying
    \begin{equation}
      \label{eq:alpha-g-estimates}
      \bqDx{\alpha_g ( L )} \leqslant \norm{g}_1 \, \qDx{L} \text{.}
    \end{equation}
    The energy-momentum transfer of $\alpha_g (L)$ is contained in
    $\supp \, \tilde{g}$.
  \end{lemma}
  \begin{proof}
    By translation invariance of the norm $\norm{~.~}$ and of the
    seminorms $\qD$ as established in
    Proposition~\ref{pro:seminorm-translation-invariance}, the
    (measurable) integrand on the right-hand side of
    \eqref{eq:alpha-g-def} is majorized by the functions $x \mapsto
    \abs{g ( x )} \, \norm{L}$ and $x \mapsto \abs{g ( x )} \,
    \qDx{L}$ for any bounded Borel set $\Delta$.  These are
    Lebesgue-integrable and, therefore, $\alpha_g ( L )$ exists as a
    unique element of the completion of $( \Lfrak , \Tfrak_q )$,
    satisfying the claimed estimates \eqref{eq:alpha-g-estimates}.

    Consider an arbitrary function $h \in L^1 \bigl( \Rsone , d^{s +
    1} x \bigr)$. According to Fubini's Theorem
    \cite[II.16.3]{fell/doran:1988a} in combination with translation
    invariance of Lebesgue measure,
    \begin{multline*}
      \int_{\Rsone} d^{s + 1} y \; h ( y ) \, \alpha_y \bigl( \alpha_g
      ( L ) \bigr) = \int_{\Rsone} d^{s + 1} y \; h ( y )
      \int_{\Rsone} d^{s + 1} x \; g ( x ) \, \alpha_{x + y} ( L ) \\ 
      = \int_{\Rsone} d^{s + 1} x \; \Bigl( \int_{\Rsone} d^{s + 1} y
      \; h ( y ) \, g ( x - y ) \Bigr) \ax ( L ) \text{,}
    \end{multline*}
    where the term in parentheses on the right-hand side is the
    convolution product $h \ast g$. Its Fourier transform is
    $\widetilde{h \ast g} ( p ) = ( 2 \pi )^{( s + 1 ) / 2} \tilde{h}
    ( p ) \tilde{g} ( p )$
    \cite[Theorem~VI.(21.41)]{hewitt/stromberg:1969} so that it
    vanishes if $\tilde{h}$ and $\tilde{g}$ have disjoint supports.
    Therefore, $\supp \tilde{h} \cap \supp \tilde{g} = \emptyset$
    entails
    \begin{equation*}
      \int_{\Rsone} d^{s + 1} y \; h ( y ) \, \alpha_y \bigl( \alpha_g
      ( L ) \bigr) = 0 \text{,}
    \end{equation*}
    demonstrating that the Fourier transform of $y \mapsto \alpha_y
    \bigl( \alpha_g ( L ) \bigr)$ has support in $\supp \tilde{g}$
    which henceforth contains the energy-momentum transfer of
    $\alpha_g ( L )$.
  \end{proof}
  Finally, it is possible to establish a property of rapid decay with
  respect to the seminorms $\qD$ for commutators of \emph{almost
    local} elements of $\Lfrak$.
  \begin{lemma}
    \label{lem:commutator-qd-decay}
    Let $L_1$ and $L_2$ belong to $\Lfrak_0$ and let $A_1$, $A_2
    \in \Afrak$ be almost local. Then
    \begin{equation*}
      \xib \mapsto \bqDx{\bcomm{\aibx ( A_1 L_1 )}{A_2 L_2}}
    \end{equation*}
    is a function that decreases faster than any power of
    $\abs{\xib}^{-1}$ when $\abs{\xib} \rightarrow \infty$.
  \end{lemma}
  \begin{proof}
    Given an approximating net $\bset{A_r \in \AOr : r > 0}$ for an
    almost local operator $A$, this can be used to construct a second
    one, $\bset{A'_r \in \AOr : r > 0}$, with $\norm{A'_r} \leqslant
    \norm{A}$ and $\norm{A -A'_r} \leqslant 2 \norm{A -A'_r}$. Nets
    with this additional property allow for an improved version of
    \eqref{eq:commutator-norm-estimate} to be used later:
    \begin{equation}
      \label{eq:special-commutator-norm-estimate}
      \bnorm{\bcomm{\aibtwox ( A )}{B}} \leqslant 2 \, \bigl(
      \norm{A - A_{\abs{\xib}}} \, \norm{B} + \norm{A} \, \norm{B -
      B_{\abs{\xib}}} \bigr) \text{,} \quad \xib \in \Rs \setminus
      \set{\zeroib} \text{.}
    \end{equation}
    
    First, we consider the special case of two elements $L_a$ and
    $L_b$ in $\Lfrak_0$ having energy-momentum transfer in compact and
    \emph{convex} subsets $\Gamma_a$ and $\Gamma_b$ of $\complement
    \fwcone$, respectively, such that $\Gamma_{a , b} \doteq (
    \Gamma_a + \Gamma_b ) - \Gamma_a$ and $\Gamma_{b , a}$, defined
    accordingly, both belong to the complement of $\fwcone$, too.
    According to Lemmas~\ref{lem:involution-invariance}
    and~\ref{lem:basic-estimate},
    \begin{multline}
      \label{eq:specialized-Ep-transfer-estimate}
      \bqDx{\bcomm{\aibx ( L_a )}{L_b}}^2 = \bpDx{\bcomm{\aibx ( L_a
      )}{L_b}^* \bcomm{\aibx ( L_a )}{L_b}} \\
      \leqslant \qDx{L_b} \, \bqDx{\aibx ( L_a )^* \bcomm{\aibx ( L_a
      )}{L_b}} + \qDx{L_a} \, \bqDx{{L_b}^* \bcomm{\aibx ( L_a
      )}{L_b}} \text{,}
    \end{multline}
    and we are left with the task to investigate $\xib \mapsto
    \bqDx{\aibx ( L_a )^* \bcomm{\aibx ( L_a )}{L_b}}$ as well as
    $\xib \mapsto \bqDx{{L_b}^* \bcomm{\aibx ( L_a )}{L_b}}$ in the
    limit of large $\abs{\xib}$. The arguments of both terms belong to
    $\Lfrak_0$ with energy-momentum transfer in the compact and convex
    sets $\Gamma_{a , b}$ and $\Gamma_{ b , a}$. Thus, relation
    \eqref{eq:commutator-integral-estimate} of
    Proposition~\ref{pro:harmonic-analysis} together with
    \eqref{eq:q-seminorm} yields for the second term
    \begin{multline}
      \label{eq:L0-commutator-estimate}
      \abs{\xib}^{2k} \bqDx{{L_b}^* \bcomm{\aibx ( L_a )}{L_b}}^2 \\
      \leqslant N ( \Delta , \Gamma_{b , a} ) \int_{\Rs} d^s y \;
      \abs{\xib}^{2k} \Bnorm{\Bcomm{\aiby \bigl({L_b}^* \bcomm{\aibx (
      L_a )}{L_b} \bigr)}{\bigl({L_b}^* \bcomm{\aibx ( L_a )}{L_b}
      \bigr)^*}} \text{.}
    \end{multline}
    Let $\bset{ L_{a , r} \in \AOr : r > 0}$ and $\bset{ L_{b , r} \in
    \AOr : r > 0}$ be approximating nets for $L_a$ and $L_b$,
    respectively, with $\norm{L_{a , r}} \leqslant \norm{L_a}$ and
    $\norm{L_{b , r}} \leqslant \norm{L_b}$; the operators ${L_{b ,
    r}}^* \bcomm{\aibx ( L_{a , r} )}{L_{b , r}} \in \Afrak (
    \Oscr_{r + \abs{\xib}} )$ then constitute the large radius part of
    approximating nets for the almost local operators ${L_b}^*
    \bcomm{\aibx ( L_a )}{L_b}$, $\xib \in \Rs$, so that for suitable
    $C_l > 0$, $l \in \Nbb$,
    \begin{equation}
      \label{eq:commutator-approximant}
      \bnorm{{L_b}^* \bcomm{\aibx ( L_a )}{L_b} - {L_{b , r}}^*
      \bcomm{\aibx ( L_{a , r} )}{L_{b , r}}} \leqslant C_l \, r^{-l}
      \text{.}
    \end{equation}
    Therefore, approximating nets $\bset{L ( a , b ; \xib )_r \in \AOr : r
    > 0}$, $\xib \in \Rs$, exist which satisfy $\norm{L ( a , b ;
    \xib )_r} \leqslant \bnorm{{L_b}^* \bcomm{\aibx ( L_a )}{L_b}}$
    and $\bnorm{{L_b}^* \bcomm{\aibx ( L_a )}{L_b} - L ( a , b ; \xib
    )_{r + \abs{\xib}}} \leqslant 2 \, C_l \, r^{-l}$, due to the
    introductory remark. This implies, according to
    \eqref{eq:special-commutator-norm-estimate}, that the integrand of
    \eqref{eq:L0-commutator-estimate} is bounded by
    \begin{multline}
      \label{eq:La-Lb-commutator-estimate}
      \abs{\xib}^{2 k} \Bnorm{\Bcomm{\aiby \bigl( {L_b}^* \bcomm{\aibx
      ( L_a )}{L_b} \bigr)}{\bigl( {L_b}^* \bcomm{\aibx ( L_a )}{L_b}
      \bigr)^*}} \\
      \mspace{-90mu} \leqslant \abs{\xib}^{2 k} 4 \, \bnorm{{L_b}^*
      \bcomm{\aibx ( L_a )}{L_b}} \bnorm{{L_b}^* \bcomm{\aibx ( L_a
      )}{L_b} - L ( a , b ; \xib )_{2^{-1} \abs{\yib}}} \\
      \leqslant
      \begin{cases}
        8 \, \abs{\xib}^{2 k} \norm{L_b}^2 \bnorm{\bcomm{\aibx ( L_a
        )}{L_b}}^2 & \text{,~} \abs{\yib} \leqslant 2 ( \abs{\xib} + 1
        ) \text{,} \\
        8 \, \norm{L_b} \abs{\xib}^{2 k} \bnorm{\bcomm{\aibx ( L_a
        )}{L_b}} C_l ( 2^{-1} \abs{\yib} - \abs{\xib} )^{-l} &
        \text{,~} \abs{\yib} > 2 ( \abs{\xib} + 1 ) \text{.}
      \end{cases}
    \end{multline}
    Having these estimates at hand, integration with respect to $\yib$
    of the right-hand side yields in both cases for $l \geqslant s +
    2$ polynomials of degree $s$ in $\abs{\xib}$ so that, due to the
    decay properties of the function $\xib \mapsto \bnorm{\bcomm{\aibx
    ( L_a )}{L_b}}$, there exists a uniform bound
    \begin{equation}
      \label{eq:qD-decay}
      \abs{\xib}^k \bqDx{{L_b}^* \bcomm{\aibx ( L_a )}{L_b}}^2
      \leqslant M \text{,} \quad \xib \in \Rs \text{.}
    \end{equation}
    The same reasoning applies to the term $\bqDx{\aibx ( L_a )^*
    \bcomm{\aibx ( L_a )}{L_b}}$, thus, by virtue of
    \eqref{eq:specialized-Ep-transfer-estimate}, establishing the
    asserted rapid decrease of the mapping $\xib \mapsto
    \bqDx{\bcomm{\aibx ( L_a )}{L_b}}$.
    
    For arbitrary almost local elements $A_1$, $ A_2 \in \Afrak$ and
    $L_1$, $L_2 \in \Lfrak_0$ one has, by use of
    Lemma~\ref{lem:basic-estimate},
    \begin{multline*}
      \bqDx{\bcomm{\aibx ( A_1 L_1 )}{A_2 L_2}} \\
      \shoveleft{\leqslant \norm{A_1} \bnorm{\bcomm{\aibx ( L_1
      )}{A_2}} \, \qDx{L_2} + \norm{A_1} \norm{A_2} \,
      \bqDx{\bcomm{\aibx ( L_1 )}{L_2}}} \\
      + \bnorm{\bcomm{\aibx ( A_1 )}{A_2}} \norm{L_2} \, \qDx{L_1} +
      \norm{A_2} \bnorm{\bcomm{\aibx ( A_1 )}{L_2}} \, \qDx{L_1}
      \text{,}
    \end{multline*}
    and rapid decay is an immediate consequence of almost locality for
    all terms but the second one on the right-hand side of this
    inequality. Using suitable decompositions of $L_1$ and $L_2$ in
    terms of elements of $\Lfrak_0$ complying pairwise with the
    special properties exploited in the previous paragraph, the
    remaining problem of decrease of the function $\xib \mapsto
    \bqDx{\bcomm{\aibx ( L_1 )}{L_2}}$ reduces to the case that has
    already been considered above.
  \end{proof}

\section{Proofs for Section~\ref{sec:particle-weights}}
  \label{sec:weights-proofs}
  
  The following results are concerned with integrability properties of
  functionals in $\Cstar$. Lemma~\ref{lem:varsigma-integrals} is an
  immediate consequence of Lemmas~\ref{lem:Poin-Bochner-integrals}
  and~\ref{lem:Lebesgue-Bochner-integrals}, whereas
  Lemma~\ref{lem:cluster-prep} prepares the proof of a kind of Cluster
  Property for \emph{positive} functionals in $\Cstar$, formulated in
  Proposition~\ref{pro:cluster}.
  \begin{lemma}
    \label{lem:varsigma-integrals}
    Let $\varsigma \in \Cstar$, $L_1$, $L_2 \in \Lfrak$ and $C
    \in \Cfrak$.
    \begin{proplist}
    \item Let $F \in L^1 \bigl( \Poin , d \mu ( \Lambda , x ) \bigr)$
      have compact support $\Ssf$, then
      \begin{subequations}
        \begin{align}
          \varsigma \bigl( {L_1}^* \alpha_F ( L_2 ) \bigr) & = \int d
          \mu ( \Lambda , x ) \; F ( \Lambda , x ) \, \varsigma \bigl(
          {L_1}^* \aLax ( L_2 ) \bigr) \text{,} \\
          \varsigma \bigl( \alpha_F ( C ) \bigr) & = \int d \mu (
          \Lambda , x ) \; F ( \Lambda , x ) \, \varsigma \bigl( \aLax
          ( C ) \bigr) \text{,}
        \end{align}
      \end{subequations}
      and there hold the estimates
      \begin{subequations}
        \begin{align}
          \babs{\varsigma \bigl( {L_1}^* \alpha_F ( L_2 ) \bigr)} &
          \leqslant \norm{F}_1 \norm{\varsigma}_\Delta \, \qDx{L_1}
          \sup_{( \Lambda , x ) \in \Ssf} \bqDx{\aLax ( L_2 )}
          \text{,} \\
          \babs{\varsigma \bigl( \alpha_F ( C ) \bigr)} & \leqslant
          \norm{F}_1 \norm{\varsigma}_\Delta \sup_{( \Lambda , x ) \in
          \Ssf} \bpDx{\aLax(C)}
        \end{align}
      \end{subequations}
      for any $\Delta$ such that $\varsigma \in \CDstar$.
    \item For any function $g \in L^1 \bigl( \Rsone , d^{s + 1} x
      \bigr)$
      \begin{equation}
        \varsigma \bigl( {L_1}^* \alpha_g ( L_2 ) \bigr) =
        \int_{\Rsone} d^{s + 1} x \; g ( x ) \, \varsigma \bigl(
        {L_1}^* \ax ( L_2 ) \bigr) \text{,}
      \end{equation}
      and a bound is given by
      \begin{equation}
        \babs{\varsigma \bigl( {L_1}^* \alpha_g ( L_2 ) \bigr)}
        \leqslant \norm{g}_1 \norm{\varsigma}_\Delta \, \qDx{L_1} \,
        \qDx{L_2}
      \end{equation}
      for any $\Delta$ with $\varsigma \in \CDstar$.
    \end{proplist}
  \end{lemma}
  \begin{lemma}
    \label{lem:cluster-prep}
    Let $L' \in \Lfrak$ and let $L \in \Lfrak$ have energy-momentum
    transfer in the compact set $\Gamma \subset \complement \fwcone$.
    If $\varsigma \in \Cstarplus$ is a positive functional belonging
    to $\CDstar$ and $\Delta'$ denotes any bounded Borel set
    containing $\Delta + \Gamma$, then
    \begin{equation}
      \label{eq:varsigmaintegral-prep}
      \int_{\Rs} d^s x \; \varsigma \bigl( L^* \aibx ( {L'}^* L' ) \,
      L \bigr) \leqslant \norm{\varsigma}_\Delta \, \qDx{L}^2
      \qDprimex{L'}^2 \text{.}
    \end{equation}
  \end{lemma}
  \begin{proof}
    Let $\Kib$ be an arbitrary compact subset of $\Rs$. Then
    \begin{equation*}
      \int_\Kib d^s x \; L^* \aibx ( {L'}^* L' ) \, L = L^* \Bigl(
      \int_\Kib d^s x \; \aibx ( {L'}^* L' ) \Bigr) L
    \end{equation*}
    belongs to the algebra of counters and exists furthermore as an
    integral in the completion of $\Cfrak$ with respect to the
    $\pD$-seminorms. Therefore, the continuous functional $\varsigma$
    can be interchanged with the integral
    \cite[Proposition~II.5.7]{fell/doran:1988a} to give
    \begin{equation*}
      \int_\Kib d^s x \; \varsigma \bigl( L^* \aibx ( {L'}^* L' ) \, L
      \bigr) = \varsigma \Bigl( L^* \int_\Kib d^s x \; \aibx ( {L'}^*
      L' ) \; L \Bigr) \text{.}
    \end{equation*}
    Making use of the positivity of $\varsigma$, an application of
    Lemma~\ref{lem:basic-estimate} leads to the estimate
    \begin{multline*}
      0 \leqslant \int_\Kib d^s x \; \varsigma \bigl( L^* \aibx (
      {L'}^* L' ) \, L \bigr) \leqslant \norm{\varsigma}_\Delta \,
      \BpDx{L^* \int_\Kib d^s x \; \aibx ( {L'}^* L' ) \; L} \\
      \leqslant \norm{\varsigma}_\Delta \, \qDx{L}^2 \Bnorm{\EDprime
      \int_\Kib d^s x \; \aibx ( {L'}^* L' ) \; \EDprime} \text{,}
    \end{multline*}
    which survives in the limit $\Kib \nearrow \Rs$. Since the
    right-hand side stays finite in this procedure the function $\xib
    \mapsto \varsigma \bigl( L^* \aibx ( {L'}^* L' ) \, L \bigr)$ is
    integrable as a consequence of the Monotone Convergence
    Theorem; its integral over $\Rs$ satisfies the asserted estimate
    due to equation \eqref{eq:q-seminorm}.
  \end{proof}

  Next comes the proof of the Cluster Property for positive
  functionals in $\Cstar$.
  \begin{proof}[Proposition~\ref{pro:cluster}]
    Commuting $A_1 L'_1$ and $\aibx ( {L_2}^* A_2 )$ in the argument
    of \eqref{eq:cluster-function}, we get the estimate
    \begin{multline}
      \label{eq:varsigma-sum-estimate}
      \babs{\varsigma \bigl( ( {L_1}^* A_1 L'_1 ) \aibx ( {L_2}^* A_2
      L'_2 ) \bigr)} \\
      \leqslant \babs{\varsigma \bigl( {L_1}^* \bcomm{A_1 L'_1}{\aibx
      ( {L_2}^* A_2 )} \aibx ( L'_2 ) \bigr)} + \babs{\varsigma
      \bigl( {L_1}^* \aibx ( {L_2}^* A_2 ) A_1 L'_1 \aibx ( L'_2 )
      \bigr)} \text{.}
    \end{multline}
    Making use of Lemma~\ref{lem:basic-estimate} and
    Proposition~\ref{pro:seminorm-translation-invariance}, the first
    term on the right-hand side turns out to be integrable over $\Rs$,
    due to almost locality of the operators encompassed by the
    commutator:
    \begin{multline}
      \label{eq:first-varsigmaintegral-estimate}
      \int_{\Rs} d^s x \; \babs{\varsigma \bigl( {L_1}^* \bcomm{A_1
      L'_1}{\aibx ( {L_2}^* A_2 )} \aibx ( L'_2 ) \bigr)} \\
      \leqslant \norm{\varsigma}_\Delta \, \qDx{L_1} \, \qDx{L'_2}
      \int_{\Rs} d^s x \; \bnorm{\bcomm{A_1 L'_1}{\aibx ( {L_2}^* A_2
      )}} \text{.}
    \end{multline}
    By positivity of $\varsigma$, application of the Cauchy-Schwarz
    inequality yields the following bound for the second term of
    \eqref{eq:varsigma-sum-estimate}:
    \begin{multline}
      \label{eq:varsigma-cauchy-schwarz}
      \varsigma \bigl( {L_1}^* \aibx ( {L_2}^* A_2 {A_2}^* L_2 ) L_1
      \bigr)^{1/2} \, \varsigma \bigl( \aibx ( {L'_2}^* ) {L'_1}^*
      {A_1}^* A_1 L'_1 \aibx ( L'_2 ) \bigr)^{1/2} \\
      = 2^{-1} \Bigl( \varsigma \bigl( {L_1}^* \aibx ( {L_2}^* A_2
      {A_2}^* L_2 ) L_1 \bigr) + \varsigma \bigl( \aibx ( {L'_2}^* )
      {L'_1}^* {A_1}^* A_1 L'_1 \aibx ( L'_2 ) \bigr) \Bigr) \text{.}
    \end{multline}
    Integration of the first term on the right-hand side is possible,
    according to Lemma~\ref{lem:cluster-prep}:
    \begin{equation}
      \label{eq:second-varsigmaintegral-estimate}
      \int_{\Rs} d^s x \; \varsigma \bigl( {L_1}^* \aibx ( {L_2}^* A_2
      {A_2}^* L_2 ) L_1 \bigr) \leqslant \norm{\varsigma}_\Delta \, 
      \qDx{L_1}^2 q_{\Delta_1} ( {A_2}^* L_2 )^2 \text{,}
    \end{equation}
    where $\Delta_1$ is any bounded Borel set containing the sum of
    $\Delta$ and the energy-momentum transfer $\Gamma_1$ of $L_1$.
    Upon commuting $\aibx ( {L'_2}^* )$ and $\aibx ( L'_2 )$ to the
    interior in the second term of \eqref{eq:varsigma-cauchy-schwarz},
    it turns out to be bounded by (cf.~Lemma~\ref{lem:basic-estimate}
    and Proposition~\ref{pro:seminorm-translation-invariance})
    \begin{multline}
      \label{eq:second-varsigma-comm-estimate}
      \babs{\varsigma \bigl( \bcomm{\aibx ( {L'_2}^* )}{{L'_1}^*}
      {A_1}^* A_1 L'_1 \aibx ( L'_2 ) \bigr)} + \babs{\varsigma \bigl(
      {L'_1}^* \aibx ( {L'_2}^* ) {A_1}^* A_1 \bcomm{L'_1}{\aibx (
      L'_2 )} \bigr)} \\ 
      \shoveright{+ \babs{\varsigma \bigl( {L'_1}^* \aibx ( {L'_2}^* )
      {A_1}^* A_1 \aibx ( L'_2 ) L'_1 \bigr)}} \\
      \leqslant \norm{A_1}^2 \Bigl( \norm{\varsigma}_\Delta \bigl(
      \norm{L'_1} \, \qDx{L'_2} + \norm{L'_2} \, \qDx{L'_1} \bigr)
      \bqDx{\bcomm{L'_1}{\aibx ( L'_2 )}} + \varsigma \bigl( {L'_1}^*
      \aibx ( {L'_2}^* L'_2 ) L'_1 \bigr) \Bigr) \text{,} \\
      ~
    \end{multline}
    where again use is made of the positivity of $\varsigma$.
    Lemma~\ref{lem:commutator-qd-decay} on rapid decay of commutators
    of almost local operators with respect to the $\qD$-seminorm and
    Lemma~\ref{lem:cluster-prep} show integrability of the right-hand
    side of \eqref{eq:second-varsigma-comm-estimate} over $\Rs$, where
    in view of \eqref{eq:varsigmaintegral-prep} the integral is
    bounded by a term proportional to $\norm{\varsigma}_\Delta$.
    Combining this result with
    \eqref{eq:first-varsigmaintegral-estimate} and
    \eqref{eq:second-varsigmaintegral-estimate} establishes the
    assertion.
  \end{proof}
  The Cluster Property has been proved under the fairly general
  assumption of almost locality of the operators involved. It also
  holds, if the mapping $\xib \mapsto \bpDx{{L_1}^* \aibx ( L_2 )}$,
  $\xib \in \Rs$, is integrable for given $L_1$, $L_2 \in \Lfrak$ and
  the continuous functional $\varsigma$ belongs to $\CDstar$.  Another
  consequence of this combination of properties concerns weakly
  convergent nets $\bset{\varsigma_\iota : \iota \in J}$ of
  functionals from bounded subsets of $\CDstar$: a kind of Dominated
  Convergence Theorem.
  \begin{lemma}
    \label{lem:varsigmanet-convergence}
    Let $L_1$, $L_2 \in \Lfrak$ be such that $\xib \mapsto
    \bpDx{{L_1}^* \aibx ( L_2 )}$ is integrable and consider the
    weakly convergent net $\bset{\varsigma_\iota : \iota \in J}$ in
    a bounded subset of $\CDstar$ with limit $\varsigma$. Then
    \begin{equation}
      \label{eq:integral-net-limits}
      \int_{\Rs} d^s x \; \varsigma \bigl( {L_1}^* \aibx ( L_2 )
      \bigr) = \lim_\iota \int_{\Rs} d^s x \; \varsigma_\iota \bigl(
      {L_1}^* \aibx ( L_2 ) \bigr) \text{.}
    \end{equation}
  \end{lemma}
  \begin{proof}
    By assumption of integrability of $\xib \mapsto \bpDx{{L_1}^*
    \aibx ( L_2 )}$, there exists a compact set $\Kib$ to any
    $\epsilon > 0$ such that
    \begin{equation}
      \label{eq:complement-integral}
      \int_{\complement \Kib} d^s x \; \bpDx{{L_1}^* \aibx ( L_2 )} <
      \epsilon \text{.}
    \end{equation}
    Moreover, by
    Proposition~\ref{pro:lcs-continuity-differentiability} and
    Corollary~\ref{cor:sesquilinear-product}, $\xib \mapsto {L_1}^*
    \aibx ( L_2 )$ is a continuous mapping on $\Rs$ with respect to
    the $\pD$-topology, hence uniformly continuous on $\Kib$. This
    means that there exists $\delta > 0$ such that $\xib$, $\xib' \in
    \Kib$ and $\abs{\xib - \xib'} < \delta$ imply
    \begin{equation}
      \label{eq:pD-delta-estimate}
      \bpDx{{L_1}^* \aibx ( L_2 ) - {L_1}^* \aibxprime ( L_2 )} <
      \epsilon \text{.}
    \end{equation}
    By compactness of $\Kib$, we can find finitely many elements
    $\xib_1$, \dots, $\xib_N \in \Kib$ so that the $\delta$-balls
    around these points cover all of $\Kib$, and, since $\varsigma$ is
    the weak limit of the net $\bset{\varsigma_\iota : \iota \in J}$,
    there exists $\iota_0 \in J$ with the property that $\iota \succ
    \iota_0$ entails
    \begin{equation}
      \label{eq:net-estimate}
      \babs{\varsigma \bigl( {L_1}^* \alpha_{\xib_i} ( L_2 ) \bigr) -
      \varsigma_\iota \bigl( {L_1}^* \alpha_{\xib_i} ( L_2 ) \bigr)} < 
      \epsilon
    \end{equation}
    for any $i=1$, \dots, $N$. Now, selecting for $\xib \in \Kib$ an
    appropriate $\xib_k$ in a distance less than $\delta$ and making
    use of \eqref{eq:pD-delta-estimate} and \eqref{eq:net-estimate},
    one has for any $\xib \in \Kib$ and $\iota \succ \iota_0$
    \begin{multline*}
      \babs{\varsigma \bigl( {L_1}^* \aibx ( L_2 ) \bigr) -
      \varsigma_\iota \bigl( {L_1}^* \aibx ( L_2 ) \bigr)} \\
      \leqslant
      \norm{\varsigma}_\Delta \, \epsilon + \babs{\varsigma \bigl(
      {L_1}^* \alpha_{\xib_k} (L_2 ) \bigr) - \varsigma_\iota \bigl(
      {L_1}^* \alpha_{\xib_k} ( L_2 ) \bigr)} +
      \norm{\varsigma_\iota}_\Delta \, \epsilon
      \leqslant \epsilon \, ( 1 + \norm{\varsigma}_\Delta +
      \norm{\varsigma_\iota}_\Delta ) \text{.}
    \end{multline*} 
    For these indices we thus arrive at the estimate
    \begin{multline*}
      \Babs{\int_{\Rs} d^s x \; \Bigl( \varsigma \bigl( {L_1}^* \aibx
      ( L_2 ) \bigr) - \varsigma_\iota \bigl( {L_1}^* \aibx ( L_2 )
      \bigr) \Bigr)} \\
      \shoveleft{\leqslant \Babs{\int_\Kib d^s x \; \Bigl( \varsigma
      \bigl( {L_1}^* \aibx ( L_2 ) \bigr) - \varsigma_\iota \bigl(
      {L_1}^* \aibx ( L_2 ) \bigr) \Bigr)}} \\
      + \Babs{\int_{\complement \Kib} d^s x \; \Bigl(
      \varsigma \bigl( {L_1}^* \aibx ( L_2 ) \bigr) - \varsigma_\iota
      \bigl( {L_1}^* \aibx ( L_2 ) \bigr) \Bigr)} \\
      \leqslant \epsilon \, ( 1 + \norm{\varsigma}_\Delta +
      \norm{\varsigma_\iota}_\Delta ) \int_\Kib d^s x + (
      \norm{\varsigma}_\Delta + \norm{\varsigma_\iota}_\Delta ) \,
      \epsilon \text{,}
    \end{multline*}
    where use is made of \eqref{eq:complement-integral}. Since the
    index $\iota_0$ can be determined appropriately for arbitrarily
    small $\epsilon$, this inequality proves the possibility to
    interchange integration and the limit with respect to $\iota$ as
    asserted in \eqref{eq:integral-net-limits}.
  \end{proof}

  The proof of the Spectral Property of functionals in $\Cstar$ relies
  on the Lemmas established above.
  \begin{proof}[Proposition~\ref{pro:spectral-property}]
    According to Lemma~\ref{lem:Lebesgue-Bochner-integrals}, for $g
    \in L^1 \bigl( \Rsone , d^{s + 1} x \bigr)$ the operator
    \begin{equation*}
      \alpha_g ( L_2 ) = \int_{\Rsone} d^{s + 1} x \; g ( x ) \,
      \ax ( L_2 )
    \end{equation*}
    lies in the completion of $( \Lfrak , \Tfrak_q )$ with
    energy-momentum transfer contained in $\supp \tilde{g}$. If
    $\varsigma$ belongs to $\CDstar$, we infer from $\supp \tilde{g}
    \subseteq \complement ( \fwcone - \Delta )$ that $\alpha_g ( L_2 )
    \ED = 0$ and henceforth, by Lemma~\ref{lem:delta-annihilation},
    $\bqDx{\alpha_g ( L_2 )} = 0$. Lemma~\ref{lem:varsigma-integrals}
    then yields
    \begin{equation*}
      \Babs{\int_{\Rsone} d^{s + 1} x \; g ( x ) \, \varsigma \bigl(
      {L_1}^* \ax ( L_2 ) \bigr)} =  \babs{\varsigma \bigl( {L_1}^*
      \alpha_g ( L_2 ) \bigr)} \leqslant \norm{\varsigma}_\Delta \,
      \qDx{L_1} \, \bqDx{\alpha_g ( L_2 )}
    \end{equation*}
    which, according to the preceding considerations, entails
    \begin{equation}
      \label{eq:spectral-support-integral}
      \int_{\Rsone} d^{s + 1} x \; g ( x ) \, \varsigma \bigl( {L_1}^*
      \ax ( L_2 ) \bigr) = 0 \text{.}
    \end{equation}
    Now, let $g' \in L^1 \bigl( \Rsone , d^{s + 1} x \bigr)$ have
    $\supp \tilde{g'} \subseteq \complement ( \fwcone - q )$ with
    $\Delta$ lying in $q -\fwcone$, then $\supp \tilde{g'} \subseteq
    \complement ( \fwcone - \Delta )$ and
    \eqref{eq:spectral-support-integral} is satisfied, thus proving
    the assertion.
  \end{proof}

  Recall that the following two proofs require the function $h$ to
  belong to $C_{0 , c} ( \Rs )$, the space of continuous functions in
  $C ( \Rs )$ which approximate a constant value in the limit
  $\abs{\vib} \rightarrow \infty$.
  \begin{proof}[Proposition~\ref{pro:translation-invariance}]
    Due to translation invariance of Lebesgue measure, one has for any
    finite time $t$ and any given $x = ( x^0 , \xib ) \in \Rsone$
    \begin{equation*}
      \rho_{h , t} \bigl( \alpha_{( x^0 , \xib )} ( C ) \bigr) = T ( t
      )^{-1} \int_{t + x^0}^{t + x^0 + T ( t )} d \tau \int_{\Rs} d^s
      y \; h \bigl( ( \tau - x^0 )^{-1} ( \yib - \xib ) \bigr) \,
      \omega \bigl( \alpha_{( \tau , \yib )} ( C ) \bigr) \text{.}
    \end{equation*}
    Accordingly, $\babs{\rho_{h , t} ( C ) - \rho_{h , t} \bigl(
    \alpha_{( x^0 , \xib )} ( C ) \bigr)}$ can be split into a sum of
    three integrals to be estimated separately:
    \begin{align*}
      \Babs{T ( t )^{-1} \int_t^{t + x^0} d \tau \int_{\Rs} d^s y \; h
      ( \tau^{-1} \yib ) \, \omega \bigl( \alpha_{( \tau , \yib )} ( C
      ) \bigr)} & \leqslant \abs{T ( t )}^{-1} \abs{x^0} \,
      \norm{h}_\infty \, \pDx{C} \text{,} \\
      \Babs{T ( t )^{-1} \int_{t + x^0 + T ( t )}^{t + T ( t )} d \tau
      \int_{\Rs} d^s y \; h ( \tau^{-1} \yib ) \, \omega \bigl(
      \alpha_{( \tau , \yib )} ( C ) \bigr)} & \leqslant \abs{T ( t
      )}^{-1} \abs{x^0} \, \norm{h}_\infty \, \pDx{C} \text{.}
    \end{align*}
    Both $\rho_{h , t} ( C )$ and $\rho_{h , t} \bigl( \alpha_{( x^0 ,
    \xib )} ( C ) \bigr)$ contribute to the third one
    \begin{multline*}
      \Babs{T ( t )^{-1} \int_{t + x^0}^{t + x^0 + T ( t )} d \tau
      \int_{\Rs} d^s y \; \bigl[ h ( \tau^{-1} \yib ) - h \bigl( (
      \tau - x^0 )^{-1} ( \yib - \xib ) \bigr) \bigr] \, \omega \bigl(
      \alpha_{( \tau , \yib )} ( C ) \bigr)} \\
      \leqslant \sup_{\tau \in I_{t , x^0}} \, \sup_{\yib \in \Rs}
      \babs{h ( \tau^{-1} \yib ) - h \bigl( ( \tau - x^0 )^{-1} ( \yib
      - \xib ) \bigr)} \, \pDx{C} \text{,}
    \end{multline*}
    where we used the abbreviation $I_{t , x^0}$ for the interval of
    $\tau$-integration. Note that for $\abs{t}$ large enough division
    by $\tau$ and $\tau - x^0$ presents no problem. We finally arrive
    at the following estimate, setting $\zib_\tau \doteq \zib + ( \tau
    - x^0 )^{-1} ( x^0 \zib - \xib )$, 
    \begin{equation}
      \label{eq:translation-invariance-estimate}
      \babs{\rho_{h , t} ( C ) - \rho_{h , t} \bigl( \alpha_{( x^0
      ,\xib )} ( C ) \bigr)} \leqslant \Bigl( 2 \, \abs{T ( t )}^{-1}
      \abs{x^0} \, \norm{h}_\infty + \sup_{\tau \in I_{t , x^0}} \,
      \sup_{\zib \in \Rs} \abs{h ( \zib ) - h ( \zib_\tau )} \Bigr) \,
      \pDx{C} \text{.}
    \end{equation}
    Since, by assumption, $h$ approaches a constant value for
    $\abs{\zib} \rightarrow \infty$, there exists to $\epsilon > 0$ a
    compact ball $\Kib$ in $\Rs$ so that $\zib \in \complement \Kib$
    implies $\abs{h ( \zib ) - h ( \zib_\tau )} < \epsilon$ for large
    $\abs{\tau}$. On the other hand, the net $\bset{\zib_\tau : \tau
    \in \Rbb}$ approximates $\zib$ uniformly on compact subsets of
    $\Rs$ in the limit $\abs{\tau} \rightarrow \infty$; as a
    consequence of continuity of $h$, i.\,e., uniform continuity on
    compacta, $\abs{h ( \zib ) - h ( \zib_\tau )} < \epsilon$ also
    holds for $\zib \in \Kib$ in this limit. Thus, for large
    $\abs{\tau}$ the term $\sup_{\zib \in \Rs} \babs{h ( \zib ) - h (
    \zib_\tau )}$ falls below any given positive bound so that the
    right-hand side of \eqref{eq:translation-invariance-estimate}
    is seen to vanish with $\abs{t} \rightarrow \infty$, since then
    $\abs{T ( t )}$ exceeds any positive value. The same holds true
    for the limit of the left-hand-side, $\babs{\sigma ( C ) - \sigma
    \bigl( \ax ( C ) \bigr)}$, which establishes the assertion.
  \end{proof}
  Last in this sequence of proofs comes that of the existence of lower
  bounds.
  \begin{proof}[Proposition~\ref{pro:lower-bounds}]
    Consider the functional $\rho_{h , t}$ defined via $\omega \in
    \Sscr ( \Delta )$ at finite time $t$. Application of the
    Cauchy-Schwarz inequality with respect to the inner product of
    square-integrable functions $f$ and $g$,
    \begin{equation*}
      ( f , g )_t \doteq T ( t )^{-1} \int_t^{t + T ( t )} d \tau \;
      \overline{f ( \tau )} \, g ( \tau ) \text{,}
    \end{equation*}
    to the absolute square of $\rho_{h , t}$ yields the estimate
    \begin{equation}
      \label{eq:csi-estimate}
      \babs{\rho_{h , t} ( C )}^2 \leqslant T ( t )^{-1} \int_t^{t + T
      ( t )} d \tau \; \Babs{\int_{\Rs} d^s x \; h ( \tau^{-1} \xib )
      \, \omega \bigl( \alpha_\tau \bigl( \aibx ( C ) \bigr) \bigr)}^2
      \text{.}
    \end{equation}
    Now, let $\Kib \subset \Rs$ be compact; due to the positivity of
    the state $\omega \in \Sscr ( \Delta )$,
    \cite[Proposition~2.3.11(b)]{bratteli/robinson:1987} and Fubini's
    Theorem \cite[II.16.3]{fell/doran:1988a} lead to the following
    estimate for arbitrary $\tau$:
    \begin{multline}
      \Babs{\omega \Bigl( \alpha_\tau \Bigl( \int_\Kib d^s x \; h (
      \tau^{-1} \xib ) \, \aibx ( C ) \Bigr) \Bigr)}^2 \\
      \leqslant \omega \Bigl( \alpha_\tau \Bigl( \int_\Kib d^s x
      \int_\Kib d^s y \; h ( \tau^{-1} \yib ) \, h ( \tau^{-1} \xib )
      \, \aiby ( C^* ) \, \aibx ( C ) \Bigr) \Bigr) \text{.}
    \end{multline}
    Commuting $\omega \circ \alpha_\tau$ and the
    integrals and passing to the limit $\Kib \nearrow \Rs$, one
    arrives, on account of the assumed integrability of the mapping $\xib
    \mapsto \bpDx{C^* \aibx ( C )}$, at
    \begin{multline}
      \label{eq:pos-omega-estimate}
      \Babs{\int_{\Rs} d^s x \; h ( \tau^{-1} \xib ) \, \omega \bigl(
      \alpha_\tau \bigl( \aibx ( C ) \bigr) \bigr)}^2 \\
      \leqslant \int_{\Rs} d^s x \int_{\Rs} d^s y \; h ( \tau^{-1}
      \yib ) \, h ( \tau^{-1} \xib ) \, \omega \bigl( \alpha_\tau
      \bigl( \aiby ( C^* ) \aibx ( C ) \bigr) \bigr) \leqslant
      \norm{h}_\infty^2 \int_{\Rs} d^s x \; \bpDx{C^* \aibx ( C )}
      \text{.} \\
      ~
    \end{multline}
    Combination of \eqref{eq:csi-estimate} and
    \eqref{eq:pos-omega-estimate} finally yields
    \begin{equation}
      \label{eq:square-rho-estimate}
      \babs{\rho_{h , t} ( C )}^2 \leqslant T ( t )^{-1} \int_t^{t + T
      ( t )} d \tau \int_{\Rs} d^s x \int_{\Rs} d^s y \; h ( \tau^{-1}
      \yib ) \, h ( \tau^{-1} \xib ) \, \omega \bigl( \alpha_\tau
      \bigl( \aiby ( C^* ) \aibx ( C ) \bigr) \bigr) \text{.}
    \end{equation}
    We want to replace the term $h ( \tau^{-1} \xib )$ by the norm
    $\norm{h}_\infty$ and, to do so, define the function $h_+ \doteq (
    \norm{h}_\infty h - h^2 )^{1/2}$ which is a non-negative element
    of $C_{0 , c} ( \Rs )$ as is $h$ itself. Then for any $\zib$,
    $\zib' \in \Rs$ there holds the equation
    \begin{equation}
      \label{eq:hh-decomposition}
      \norm{h}_\infty h ( \zib ) = h ( \zib ) \, h ( \zib' ) + h_+ (
      \zib ) \, h_+ ( \zib' ) +  h_+ ( \zib ) \bigl( h_+ ( \zib ) -
      h_+ ( \zib' ) \bigr) + h ( \zib ) \bigl( h ( \zib ) - h ( \zib'
      ) \bigr) \text{.}
    \end{equation}
    Next, for an arbitrary function $g \in C_{0 , c} ( \Rs )$ the
    following inequality can be based on an application of Fubini's
    Theorem and the reasoning of \eqref{eq:rho-estimate}: 
    \begin{multline}
      \label{eq:intermediate-integral-estimate}
      \Babs{T ( t )^{-1} \int_t^{t + T ( t )} \negthickspace
      \negthickspace d \tau \int_{\Rs} \negthinspace d^s x
      \int_{\Rs} \negthinspace d^s y \; g ( \tau^{-1} \yib ) \bigl( g
      ( \tau^{-1} \yib ) - g ( \tau^{-1} \xib ) \bigr) \, \omega
      \bigl( \alpha_\tau \bigl( \aiby ( C^* ) \aibx ( C ) \bigr)
      \bigr)} \\
      = \Babs{\int_{\Rs} d^s x \; T ( t )^{-1} \int_t^{t + T ( t )} d
      \tau \int_{\Rs} d^s z \; \tau^s g ( \zib ) \bigl( g ( \zib ) - g
      \bigl( \zib_\tau ( \xib ) \bigr) \bigr) \, \omega \bigl(
      \alpha_{( \tau , \tau \zib )} \bigl( C^* \aibx ( C ) \bigr)
      \bigr)} \\
      \leqslant \norm{g}_\infty \int_{\Rs} d^s x \; \sup_{\tau \in
      I_t} \, \sup_{\zib \in \Rs} \babs{g ( \zib ) - g \bigl(
      \zib_\tau ( \xib ) \bigr)} \, \bpDx{C^* \aibx ( C )} \text{.}
    \end{multline}
    Here we use the coordinate transformation $\xib \rightsquigarrow
    \xib + \yib$ followed by the transformation $\yib \rightsquigarrow
    \zib \doteq \tau^{-1} \yib$ and introduce the abbreviations
    $\zib_\tau ( \xib ) \doteq \tau^{-1} \xib + \zib$ as well as $I_t$
    for the interval of $\tau$-integration. Similar to the proof of
    Proposition~\ref{pro:translation-invariance}, the expression
    $\sup_{\tau \in I_t} \sup_{\zib \in \Rs} \abs{g ( \zib ) - g
      \bigl( \zib_\tau ( \xib ) \bigr)}$ is seen to vanish for all
    $\xib \in \Rs$ in the limit of large $\abs{t}$ so that by
    Lebesgue's Dominated Convergence Theorem the left-hand side of
    \eqref{eq:intermediate-integral-estimate} converges to 0. This
    reasoning in particular applies to the functions $h$ as well as
    $h_+$ and thus to the third and fourth term on the right of
    equation \eqref{eq:hh-decomposition}. On the other hand,
    substitution of $h$ by $h_+$ in the integral
    \eqref{eq:square-rho-estimate} likewise gives a non-negative
    result for all times $t$. Combining all this information and
    specializing to a subnet $\bset{t_\iota : \iota \in J}$
    approximating $+ \infty$ or $- \infty$, one arrives at the
    following asymptotic version of \eqref{eq:square-rho-estimate}:
    \begin{multline*}
      \lim_\iota \babs{\rho_{h , t_\iota} ( C )}^2 \\
      \leqslant \lim_\iota \norm{h}_\infty T ( t_\iota)^{-1}
      \int_{t_\iota}^{t_\iota + T( t_\iota )} d \tau \int_{\Rs} d^s x
      \int_{\Rs} d^s y \; h ( \tau^{-1} \yib ) \, \omega \bigl(
      \alpha_\tau \bigl( \aiby ( C^* ) \aibx ( C ) \bigr) \bigr) \\
      = \norm{h}_\infty \lim_\iota \int_{\Rs} d^s x \; \rho_{h ,
      t_\iota} \bigl( C^* \aibx ( C ) \bigr) \text{.}
    \end{multline*}
    By Lemma~\ref{lem:varsigmanet-convergence}, this result extends to
    the limit functional $\sigma$, thus yielding
    \eqref{eq:lower-bounds}.
  \end{proof}

  The equivalent characterization of particle weights in
  Theorem~\ref{the:particle-weight} is immediate apart from an
  application of Stone's Theorem. 
  \begin{proof}[Theorem~\ref{the:particle-weight}]
    \begin{Prooflist}
    \item The various properties stated in the Theorem are readily
      established, once the GNS-construction has been carried out.
      The existence of a strongly continuous unitary representation of
      spacetime translations in $( \pi_w , \Hscr_w )$ is a direct
      consequence of translation invariance of the particle weight
      $\scp{~.~}{~.~}$ and its continuity under Poincar\'e
      transformations with respect to $q_w$. Stone's Theorem
      (cf.~\cite[Chapter~6, \S\,2]{barut/raczka:1980} and
      \cite[Theorem~VIII.(33.8)]{hewitt/ross:1970}) connects the
      spectrum of the generator $P_w = ( P_w^{\, \mu} )$ of the
      unitary representation with the support of the Fourier transform
      of $x \mapsto \bscp{L_1}{\ax ( L_2 )}$, by virtue of the
      relation
      \begin{multline}
        \label{eq:stones-theorem}
        \int_{\Rsone} d^{s + 1} x \; g ( x ) \, \bscp{L_1}{\ax ( L_2
        )} = \int_{\Rsone} d^{s + 1} x \; g ( x ) \, \bscpx{L_1}{U_w (
        x )}{L_2} \\
        = ( 2 \pi )^{( s + 1 )/2} \bscpx{L_1}{\tilde{g} ( P_w )}{L_2}
      \end{multline}
      which holds for any $L_1$, $L_2 \in \Lfrak$ and any $g \in L^1
      \bigl( \Rsone , d^{s + 1} x \bigr)$. To clarify this fact, note,
      that the projection-valued measure $E_w (~.~)$ corresponding to
      $P_w$ is regular, which means that for each Borel set $\Delta'$
      the associated projection $\EwDprime$ is the strong limit of the
      net $\bset{E_w ( \Gamma' ) : \Gamma' \subseteq \Delta' \;
      \text{compact}}$.  For each compact $\Gamma \subseteq
      \complement ( \fwcone - q )$ consider an infinitely often
      differentiable function $\tilde{g}_\Gamma$ with support in the
      set $\complement ( \fwcone - q )$ enveloping the characteristic
      function for $\Gamma$ \cite[Satz~7.7]{jantscher:1971}: $0
      \leqslant \chi_\Gamma \leqslant \tilde{g}_\Gamma$. According to
      the assumption of Definition~\ref{def:particle-weight}, the
      left-hand side of \eqref{eq:stones-theorem} vanishes for any
      $g_\Gamma$ of the above kind, and this means that all the
      bounded operators $\tilde{g}_\Gamma ( P_w )$ equal $0$ not only
      on the dense subspace spanned by vectors $\ket{L}$, $L \in
      \Lfrak$, but on all of $\Hscr_w$. Due to the fact that
      $\tilde{g}_\Gamma$ majorizes $\chi_\Gamma$, this in turn implies
      $\chi_\Gamma ( P_w ) = \EwGamma = 0$ and thus, by arbitrariness
      of $\Gamma \subseteq \complement ( \fwcone - q )$ in connection
      with regularity, the desired relation $E_w \bigl( \complement (
      \fwcone - q ) \bigr)= 0$.
    \item The reversion of the above arguments in order to establish
      that the scalar product on $\Hscr_w$ possesses the
      characteristics of a particle weight is self-evident.
    \end{Prooflist}
    \renewcommand{\qed}{}
  \end{proof}

  The following analogue of Lemmas~\ref{lem:Poin-Bochner-integrals}
  and \ref{lem:Lebesgue-Bochner-integrals} in terms of the
  $q_w$-topology is of importance not only for the results of
  Section~\ref{sec:particle-weights}, but also plays an important role
  in the constructions that underlie the theory of disintegration.
  \begin{lemma}
    \label{lem:alpha-ket-integral}
    Let $L \in \Lfrak$ and let $\scp{~.~}{~.~}$ be a particle weight.
    \begin{proplist}
    \item Let $F \in L^1 \bigl( \Poin , d \mu ( \Lambda , x ) \bigr)$
      have compact support $\Ssf$, then the Bochner integral
      \begin{subequations}
        \begin{equation}
          \label{eq:alpha-F-weight-integral}
          \alpha_F ( L ) = \int d \mu ( \Lambda , x ) \; F ( \Lambda
          , x ) \, \aLax ( L ) 
        \end{equation}
        lies in the completion of $\Lfrak$ with respect to the locally
        convex topology induced on it by the initial norm $\norm{~.~}$
        and the $q_w$-seminorm defined by the particle weight.
        Moreover, $\bket{\alpha_F ( L )}$ is a vector in the
        corresponding Hilbert space $\Hscr_w$ that can be written
        \begin{equation}
          \label{eq:F-ket-integral}
          \bket{\alpha_F ( L )} = \int d \mu ( \Lambda , x ) \; F (
          \Lambda , x ) \, \bket{\aLax ( L )} \text{,} 
        \end{equation}
      \end{subequations}
      its norm being bounded by $\norm{F}_1 \sup_{( \Lambda , x ) \in
      \Ssf} \bnorm{\bket{\aLax ( L )}}$.
    \item For any function $g \in L^1 \bigl( \Rsone , d^{s + 1} x
      \bigr)$ the Bochner integral
      \begin{subequations}
        \begin{equation}
          \label{eq:alpha-g-weight-integral}
          \alpha_g ( L ) = \int_{\Rsone} d^{s + 1} x \; g ( x ) \,
          \ax ( L ) 
        \end{equation}
        likewise lies in the completion of $\Lfrak$ with respect to
        the locally convex topology mentioned above. $\bket{\alpha_g (
        L )}$ is a vector in the Hilbert space $\Hscr_w$ subject to
        the relation
        \begin{equation}
          \label{eq:g-ket-integral}
          \bket{\alpha_g ( L )} = \int_{\Rsone} d^{s + 1} x \; g ( x
          ) \, \bket{\ax ( L )} = ( 2 \pi )^{( s + 1 )/2} \tilde{g}
          ( P_w ) \ket{L}
        \end{equation}
      \end{subequations}
      with norm bounded by $\norm{g}_1 \, \norm{\ket{L}}$. Here $P_w =
      ( P_w^{\, \mu} )$ denotes the generator of the unitary
      representation of spacetime translations in $( \pi_w , \Hscr_w
      )$.
    \end{proplist}
  \end{lemma}
  \begin{proof}
    \begin{prooflist}
    \item The seminorm $\qw$ induced on $\Lfrak$ by the particle
      weight is continuous with respect to Poincar\'e transformations
      so that the integrand of \eqref{eq:alpha-F-weight-integral} can
      be estimated by the Lebesgue-integrable function $( \Lambda , x
      ) \mapsto \abs{F ( \Lambda , x )} \cdot \sup_{( \Lambda , x )
      \in \Ssf} \qwx{\aLax ( L )}$. Therefore, the integral in
      question indeed exists in the completion of the locally convex
      space $\Lfrak$ not only with respect to the norm topology but
      also with respect to the seminorm $\qw$. Now, $\norm{\ket{L'}}$
      coincides with $\qwx{L'}$ for any $L' \in \Lfrak$, a relation
      which extends to the respective completions \cite[Chapter~One,
      \S\,5\,4.(4)]{koethe:1983} thus resulting in
      \eqref{eq:F-ket-integral} and the given bound for this integral.
    \item Invariance of $\scp{~.~}{~.~}$ with respect to spacetime
      translations implies translation invariance of the seminorm
      $\qw$. Therefore the integrand of
      \eqref{eq:alpha-g-weight-integral} is majorized by the
      Lebesgue-integrable function $x \mapsto \abs{g ( x )} \,
      \qwx{L}$ so that the respective integral exists in the
      completion of $\Lfrak$. The first equation of
      \eqref{eq:g-ket-integral} and its norm bound arise from the
      arguments that were already applied in the first part, whereas
      the second one is a consequence of Stone's Theorem
      (cf.~\eqref{eq:stones-theorem}).
    \end{prooflist}
    \renewcommand{\qed}{}
  \end{proof}
  With this result, we are in the position to prove
  Proposition~\ref{pro:spectral-subspace} on spectral subspaces of
  $\Hscr_w$.
  \begin{proof}[Proposition~\ref{pro:spectral-subspace}]
    The energy-momentum transfer of $A \in \Afrak$ can be described by
    the support of the Fourier transform of $x \mapsto \ax ( A )$
    considered as an operator-valued distribution. Thus, by
    assumption, for any Lebesgue-integrable function $g$ satisfying
    $\supp \tilde{g} \cap \Delta' = \emptyset$ we have $\alpha_g ( L )
    = 0$ and, by virtue of Lemma~\ref{lem:alpha-ket-integral},
    \begin{equation}
      \label{eq:g-ket-integral-vanished}
      \int_{\Rsone} d^{s + 1} x \; g ( x ) \, \bket{\ax ( L )} =
      \bket{\alpha_g ( L )} = 0 \text{.}
    \end{equation}
    Upon insertion of \eqref{eq:g-ket-integral-vanished} into the
    formulation \eqref{eq:stones-theorem} of Stone's Theorem, the
    reasoning applied in the proof of the first part of
    Theorem~\ref{the:particle-weight} yields the assertion.
  \end{proof}

  The proofs of Proposition~\ref{pro:weights-cluster} and
  Lemma~\ref{lem:Delta-bound} require considerably more work than
  the last ones.
  \begin{proof}[Proposition~\ref{pro:weights-cluster}]
    To establish this result we follow the strategy of the proof of
    Proposition~\ref{pro:cluster}. Applied to the problem at hand,
    expressed in terms of $( \pi_w , \Hscr_w )$, this yields for any
    $\xib \in \Rs$
    \begin{multline}
      \label{eq:weight-cluster-estimate1}
      \babs{\bscpx{{L_1}^* A_1 L'_1}{U_w ( \xib )}{{L_2}^* A_2 L'_2}}
      \\ 
      \leqslant \babs{\bscpx{L'_1}{\pi_w \bigl( \bcomm{{A_1}^*
      L_1}{\aibx ( {L_2}^* A_2 )} \bigr) U_w ( \xib )}{L'_2}} +
      \babs{\bscpx{L'_1}{\pi_w \bigl( \aibx ( {L_2}^* A_2 ) {A_1}^*
      L_1 \bigr) U_w ( \xib )}{L'_2}} \text{.} \\
      ~ 
    \end{multline}
    The first term on the right-hand side is majorized by the product
    of norms of its constituents $\bnorm{\bcomm{{A_1}^* L_1}{\aibx (
    {L_2}^* A_2 )}} \, \norm{\ket{L'_1}} \, \norm{\ket{L'_2}}$ as the
    particle weight is translation invariant and the representation is
    continuous. The operators involved are almost local without
    exception, so the norm of the commutator decreases rapidly,
    rendering this term integrable. The second term requires a closer
    inspection:
    \begin{multline}
      \label{eq:weight-cluster-estimate2}
      \babs{\bscpx{L'_1}{\pi_w \bigl( \aibx ( {L_2}^* A_2 ) {A_1}^*
      L_1 \bigr) U_w ( \xib )}{L'_2}} \\
      \mspace{-120mu} \leqslant \bnorm{\pi_w \bigl( \aibx ( {A_2}^*
      L_2 ) \bigr) \ket{L'_1}} \, \bnorm{\pi_w \bigl( {A_1}^* L_1
      \bigr) U_w ( \xib ) \ket{L'_2}} \\
      \leqslant 2^{- 1} \Bigl( \bnorm{\pi_w \bigl( \aibx ( {A_2}^* L_2
      ) \bigr) \ket{L'_1}}^2 + \bnorm{\pi_w \bigl( \aminusibx (
      {A_1}^* L_1 ) \bigr) \ket{L'_2}}^2 \Bigr) \text{.}
    \end{multline}
    Now, $\pi_w ( A' )$ has the same energy-momentum transfer with
    respect to the unitary representation $x \mapsto U_w ( x )$ as the
    operator $A' \in \Afrak$ has regarding the underlying positive
    energy representation, and, according to
    Proposition~\ref{pro:spectral-subspace}, $\ket{L'_i} = E_w (
    \Gamma'_i ) \ket{L'_i}$, $i = 1$, $2$. Since the spectrum of $x
    \mapsto U_w ( x )$ is restricted to a displaced forward light
    cone, all of the arguments given in the proofs of
    Propositions~\ref{pro:harmonic-analysis} and
    \ref{pro:counter-integrals} also apply to the representation
    $(\pi_w , \Hscr_w )$ so that, e.\,g., the integral
    \begin{equation*}
      \int_{\Rs} d^s x \; E_w ( \Gamma'_1 ) \pi_w \bigl( \aibx (
      {L_2}^* A_2 \, {A_2}^* L_2 ) \bigr) E_w ( \Gamma'_1 )
    \end{equation*}
    is seen to exist in the $\sw$-topology on $\BHw$. Thus
    \begin{multline}
      \label{eq:weight-cluster-estimate3}
      \int_{\Rs} d^s x \; \bnorm{\pi_w \bigl( \aibx ( {A_2}^* L_2 )
      \bigr) E_w ( \Gamma'_1 ) \ket{L'_1}}^2 \\ 
      = \int_{\Rs} d^s x \; \bscpx{L'_1}{E_w ( \Gamma'_1 ) \pi_w
      \bigl( \aibx ( {L_2}^* A_2 \, {A_2}^* L_2 ) \bigr) E_w (
      \Gamma'_1 )}{L'_1} < \infty \text{.}
    \end{multline}
    The same holds true for the other term on the right-hand side of
    \eqref{eq:weight-cluster-estimate2}, so its left-hand side is seen
    to be an integrable function of $\xib$, too. Altogether, we have
    thus established the Cluster Property for particle weights.
  \end{proof}
  \begin{proof}[Lemma~\ref{lem:Delta-bound}]
    Let $( \pi_\sigma , \Hscr_\sigma )$ denote the GNS-representation
    corresponding to the functional $\sigma \in \CDstar$ and let
    $E_\sigma (~.~)$ be the spectral measure associated with the
    generator $P_\sigma = ( P_\sigma^\mu )$ of the intrinsic
    representation of spacetime translations. For the time being,
    suppose that $\Delta'$ is an \emph{open} bounded Borel set in
    $\Rsone$. Let furthermore $L$ be an arbitrary element of $\Lfrak$
    and $A \in \Afrak$. We are interested in an estimate of the term
    $\scpx{L}{\EsDprime \pi_\sigma ( A ) \EsDprime}{L}_\sigma$. The
    spectral measure is regular, so $\EsDprime$ is the strong limit of
    the net $\set{\EsGamma : \Gamma \subset \Delta' \thickspace
    \text{compact}}$. As $\Delta'$ is assumed to be open, there
    exists for each compact subset $\Gamma$ of $\Delta'$ an infinitely
    often differentiable function $\tilde{g}_\Gamma$ with $\supp
    \tilde{g}_\Gamma \subset \Delta'$ that fits between the
    corresponding characteristic functions
    \cite[Satz~7.7]{jantscher:1971}: $\chi_\Gamma \leqslant
    \tilde{g}_\Gamma \leqslant \chi_{\Delta'}$. Thus
    \begin{equation*}
      0 \leqslant \bigl( \EsDprime - \tilde{g}_\Gamma ( P_\sigma )
      \bigr)^2 \leqslant \bigl( \EsDprime - \EsGamma \bigr)^2 \text{,}
    \end{equation*}
    from which we infer that for arbitrary $L' \in \Lfrak$
    \begin{equation}
      \label{eq:regular-Esigma-approx}
      0 \leqslant \bnorm{\bigl( \EsDprime - \tilde{g}_\Gamma (
      P_\sigma ) \bigr) \ket{L'}}^2 \leqslant \bnorm{\bigl( \EsDprime
      - \EsGamma \bigr) \ket{L'}}^2 \xrightarrow[\Gamma \nearrow
      \Delta']{} 0 \text{.}
    \end{equation}
    By density of all vectors $\ket{L'}$ in $\Hscr_\sigma$, this means
    that $\EsDprime$ is the strong limit of the net
    $\bset{\tilde{g}_\Gamma ( P_\sigma ) : \Gamma \subset \Delta'
     thickspace \text{compact}}$ for $\Gamma \nearrow \Delta'$ and
    \begin{equation}
      \label{eq:regular-approximation}
      \scpx{L}{\EsDprime \pi_\sigma ( A ) \EsDprime}{L}_\sigma =
      \lim_{\Gamma \nearrow \Delta'} \scpx{L}{\tilde{g}_\Gamma (
      P_\sigma ) \pi_\sigma ( A ) \tilde{g}_\Gamma ( P_\sigma )}{L}
      \text{.}
    \end{equation}
    The Fourier transform $\tilde{g}_\Gamma$ of the rapidly decreasing
    function $g_\Gamma$ belongs to $L^1 \bigl( \Rsone , d^{s
    + 1} x \bigr)$, so Lemma~\ref{lem:alpha-ket-integral} yields for
    the right-hand side of \eqref{eq:regular-approximation}
    \begin{multline}
      \label{eq:alpha-g-reformulation}
      \scpx{L}{\tilde{g}_\Gamma ( P_\sigma ) \pi_\sigma ( A )
      \tilde{g}_\Gamma ( P_\sigma )}{L} = ( 2 \pi )^{- ( s + 1 )} 
      \scpx{\agGamma ( L )}{\pi_\sigma ( A )}{\agGamma ( L )}_\sigma
      \\
      = ( 2 \pi )^{- ( s + 1 )} \sigma \bigl( \agGamma ( L )^* A
      \agGamma ( L ) \bigr) \text{,}
    \end{multline}
    where in the last equation we use the continuous extension of
    $\sigma$ to the argument at hand
    (cf.~Lemmas~\ref{lem:Lebesgue-Bochner-integrals} and
    \ref{lem:basic-estimate} in connection with
    Corollary~\ref{cor:sesquilinear-product}). The approximating
    functionals $\rho_{h , t}$ for $\sigma$ in the form
    \eqref{eq:rho-alt-form} with a non-negative function $h \in
    L^\infty ( \Rs , d^s x )$ satisfy the following estimate of their
    integrand by an application of
    \cite[Proposition~2.3.11]{bratteli/robinson:1987}:
    \begin{multline*}
      \babs{h ( \tau^{-1} \xib ) \, \omega \bigl( U ( \tau ) \ED \aibx
      ( \agGamma ( L )^* A \agGamma ( L ) ) \ED U ( \tau )^* \bigr)}
      \\ 
      = h ( \tau^{-1} \xib ) \, \babs{\omega \bigl( U ( \tau ) \ED
      \aibx (\agGamma ( L )^* ) \EDbar \aibx ( A ) \EDbar \aibx (
      \agGamma ( L ) ) \ED U ( \tau )^* \bigr)} \\
      \leqslant \norm{\EDbar A \EDbar} \; h ( \tau^{-1} \xib ) \,
      \omega \bigl( U ( \tau ) \ED \aibx ( \agGamma ( L )^* \agGamma (
      L ) ) \ED U ( \tau )^* \bigr) \text{.}
    \end{multline*}
    The spectral projections $\EDbar$ pertaining to the bounded, open
    Borel set $\Deltabar = \Delta + \Delta'$ can be inserted here,
    since, according to Lemma~\ref{lem:Lebesgue-Bochner-integrals},
    the energy-momentum transfer of $\agGamma ( L )$ is contained in
    $\Delta'$ by construction. An immediate consequence is
    \begin{equation*}
      \babs{\rho_{h , t} \bigl( \agGamma ( L )^* A \agGamma ( L )
      \bigr)} \leqslant \norm{\EDbar A \EDbar} \; \rho_{h , t}
      \bigl( \agGamma ( L )^* \agGamma ( L ) \bigr) \text{,}
    \end{equation*}
    which extends to the limit functional $\sigma$
    \begin{equation}
      \label{eq:Delta-bound-estimate1}
      \babs{\sigma \bigl( \agGamma ( L )^* A \agGamma ( L ) \bigr)}
      \leqslant \norm{\EDbar A \EDbar} \; \sigma \bigl( \agGamma ( L
      )^* \agGamma ( L ) \bigr) \text{.}
    \end{equation}
    Insertion of this result into \eqref{eq:alpha-g-reformulation} and
    passing to the limit $\Gamma \nearrow \Delta'$ in
    \eqref{eq:regular-approximation} yields
    \begin{equation}
      \label{eq:Delta-bound-estimate3}
      \babs{\scpx{L}{\EsDprime \pi_\sigma ( A ) \EsDprime}{L}_\sigma}
      \leqslant \norm{\EDbar A \EDbar} \,
      \scpx{L}{\EsDprime}{L}_\sigma \leqslant \norm{\EDbar A \EDbar}
      \, \scp{L}{L}_\sigma \text{.}
    \end{equation}
    Taking the supremum with respect to all $L \in \Lfrak$ with
    $\norm{\ket{L}_\sigma} \leqslant 1$ (these constitute a dense
    subset of the unit ball in $\Hscr_\sigma$), we get, through an
    application of \cite[Satz~4.4]{weidmann:1976},
    \begin{equation}
      \label{eq:Delta-final-bound1}
      \norm{\EsDprime \pi_\sigma ( A ) \EsDprime} \leqslant 2 \cdot
      \norm{\EDbar A \EDbar} \text{.}
    \end{equation}
    This establishes the defining condition
    \eqref{eq:Delta-boundedness} for $\Delta$-boundedness with $c = 2$
    in the case of an \emph{open} bounded Borel set $\Delta'$. But
    this is not an essential restriction, since an arbitrary bounded
    Borel set $\Delta'$ is contained in the open set $\Delta'_\eta$,
    $\eta > 0$, consisting of all those points $p \in \Rsone$ with
    $\inf_{p' \in \Delta'} \abs{p - p'} < \eta$. Since $\Delta'_\eta$
    is likewise a bounded Borel set, we get
    \begin{equation}
      \label{eq:Delta-final-bound2}
      \norm{\EsDprime \pi_\sigma ( A ) \EsDprime} \leqslant
      \norm{\EsDprimeeta \pi_\sigma ( A ) \EsDprimeeta} \leqslant 2
      \cdot \norm{\EDetabar A \EDetabar}
    \end{equation}
    as an immediate consequence of \eqref{eq:Delta-final-bound1} with
    $\Deltaetabar \doteq \Delta + \Delta'_\eta$. This covers the
    general case and thereby proves $\Delta$-boundedness for positive
    asymptotic functionals $\sigma \in \CDstar$.
  \end{proof}

\section{Conclusions}
  \label{sec:conclusions}
  
  The present article is based on the general point of view that the
  concept of \emph{particles} is asymptotic in nature and
  simultaneously has to be founded by making appropriate use of
  locality. This reflects the conviction that the long-standing
  problem of \emph{asymptotic completeness} of quantum field theory,
  i.\,e., the question if a quantum field theoretic model can be
  interpreted completely in terms of particles, has to be tackled by
  the aid of further restrictions on the general structure, which
  essentially are of a local character. The question is which local
  structure of a theory is appropriate in order that it governs
  scattering processes in such a way that asymptotically the physical
  states appear to clot in terms of certain entities named particles.
  The compactness and nuclearity conditions discussed in
  \cite{buchholz/porrmann:1990} and the references therein are
  examples of this kind of approach. It is not claimed that they
  already give a complete answer, but that they indicate the right
  direction.
  
  Asymptotic functionals on a certain algebra of detectors have been
  constructed that give rise to particle weights which are to be
  interpreted as mixtures of particle states. Their disintegration
  (presented in a forthcoming paper) constitutes the basis for the
  definition of mass and spin even in the case of charged states
  (cf.~\cite{buchholz/porrmann/stein:1991}). The technical problems
  arising in the course of these investigations, e.\,g., that of
  convergence in connection with Theorem~\ref{the:singular-limits} and
  those to be encountered when establishing the disintegration theory,
  might be solvable with additional information at hand that could be
  provided by the investigation of concrete models. Quantum
  electrodynamics is an example \cite{fredenhagen/freund}. So far only
  \emph{single-particle} weights have been considered. Another field
  of future research should be the inspection of coincidence
  arrangements of detectors as in \cite{araki/haag:1967}. In this
  respect, too, the analysis of concrete models might prove helpful.

  \subsection*{Acknowledgements}

    The results presented above have been worked out by Detlev
    Buchholz and the author on the basis of ideas of Buchholz. I
    should like to thank him for this opportunity of collaboration.
    Bernd Kuckert has given valuable advice in editing the final
    version of the manuscript, his help is gratefully acknowledged.
    Also acknowledged is financial support by Deutsche
    Forschungsgemeinschaft which I obtained from the Graduiertenkolleg
    ``Theoretische Elementarteilchenphysik'' at the II.~Institut f\"ur
    Theoretische Physik of the University of Hamburg.

\providecommand{\SortNoop}[1]{}

\end{document}